\documentclass[10pt,aps,pra,showpacs,preprintnumbers,twoside,twocolumn,groupedaddress,floatfix]{revtex4-1}
\usepackage{graphicx}
\usepackage{graphics}
\usepackage{color}
\usepackage{amsmath}

\usepackage{amssymb}
\usepackage{amsfonts}
\usepackage{bbm}
\usepackage[plainpages=false]{hyperref}
\hypersetup{
  pdfauthor={Heike Schwager and J. Ignacio Cirac and Geza Giedke},
  pdftitle={Dissipative spin chains: Implementation with cold atoms and steady-state properties}
  pdfsubject={},
  urlcolor=blue,
}
\newcommand{\id}{\ensuremath{\mathbbm{1}}} 
\def\ket#1{\left|#1\right>}
\def\bra#1{\left<#1\right|}
\newcommand{\ketbra}[2]{\ket{#1}\hspace{-4pt}\bra{#2}}

\begin{document}

\title{Dissipative spin chains: Implementation with cold atoms and steady-state properties}
\author{Heike Schwager$^1$, J. Ignacio Cirac$^1$, and G\'eza Giedke$^{1,2}$}
\affiliation{(1) Max--Planck--Institut f\"{u}r Quantenoptik,
Hans-Kopfermann--Str. 1, D--85748 Garching, Germany }
\affiliation{(2) M5, Zentrum Mathematik, TU M\"unchen,
  L.-Boltzmannstr. 3, D--85748 Garching, Germany}
\date{Aug 3, 2012}

\begin{abstract}
  We propose a quantum optical implementation of a class of dissipative spin
  systems, including the XXZ and Ising model, with ultra-cold atoms in optical
  lattices. Employing the motional degree of freedom of the atoms and detuned
  Raman transitions we show how to obtain engineerable dissipation and a
  tunable transversal magnetic field, enabling the study of the dynamics and
  steady-states of dissipative spin models. \\
  As an example of effects made accessible this way, we
    consider small spin chains and weak dissipation and show by numerical
    simulation that steady-state expectation values display pronounced peaks
    at certain critical system parameters. We show that this effect is
  related to degeneracies in the Hamiltonian and derive a sufficient condition
  for its occurrence.
\end{abstract}

\maketitle
  
\section{Introduction}

Quantum spin models play a fundamental role for the theoretical and
experimental study of quantum many-body effects. They represent
paradigmatic systems exhibiting, e.g., quantum phase transitions
and peculiar forms of matter \cite{Sac99}. They also provide toy
models for description of many solid state systems. Ultra-cold atoms
in optical lattices \cite{Bloch2005} have emerged as a system that
is especially suited to study the low-energy sector of quantum spin
systems with the promise to eventually simulate theoretical models
in large, controlled quantum systems.

To observe these effects, coupling to uncontrolled degrees of
freedom has to be kept to a minimum, since it leads to dissipation
and decoherence \cite{Zurek2003,BaNa11} which can mask or destroy
the quantum effects. But in recent years, it has been shown how the
coupling to an environment can be harnessed to generate useful
quantum states \cite{Beige2000,Benatti2003,Kraus2008,Muschik2010,
Verstraete2009} or perform quantum information tasks
\cite{Verstraete2009,Vollbrecht2011}. Moreover, the study of the
phase diagram of open quantum systems has turned into a fruitful
direction itself
\cite{Carmichael1980,Dimer2007,MoPa07,Diehl2008,EiPr11,Mueller2012}.

Our aim in the present work is twofold: In the first part of the paper, we
propose a scheme to realize a quantum spin system using ultra-cold atoms in an
optical lattice in which both coherent interaction and dissipation can be
engineered and controlled, enabling the study the non-equilibrium and
steady-state physics of open and driven spin systems. In the second part, we
highlight a peculiar feature of the steady-state diagram for small spin
chains: in the limit of weak dissipation, abrupt changes of steady-state
expectation values for certain critical values of the system parameters are
observed. We explain this feature and relate it to degeneracy properties of
the system Hamiltonian and derive a sufficient condition for the occurrence of
sharp peaks at critical system parameters.

\section{Physical implementation of a one-dimensional spin chain under dissipation}

Ultra-cold bosonic atoms in optical lattices are ideal candidates to
simulate spin Hamiltonians. Different theoretical and experimental
approaches \cite{Lew12} have been employed to simulate quantum spin
chains in optical lattices, for example by optical driving of two
hyperfine levels of cold bosons in the Hubbard regime
\cite{GaIg2003}. Recently, a one-dimensional chain of interacting
Ising spins has been implemented experimentally using a
Mott-Insulator of spinless bosons in a tilted optical lattice
\cite{Gr2011}.

In the following, we show theoretically how to add engineered dissipation to
the toolbox of these systems \cite{JaZo2005,CaI2011}. Specifically, we show
how to implement a system with the following properties: (i) dissipative
dynamics of Lindblad form, (ii) a tunable magnetic field in $x$-direction and
(iii) an effective spin Hamiltonian such as, e.g., the XXZ, Heisenberg or
Ising model. In the next Subsections, we first introduce the setup and explain
qualitatively how such a one-dimensional spin chain in a tunable magnetic
field under engineerable dissipation can be realized with cold atoms in
optical lattices. In the subsequent Subsections we give specific requirements
and parameters and details of the derivation for (i)-(iii).

\subsection{Setup and qualitative description}\label{setup}
\begin{figure}
 \includegraphics[scale=0.6]{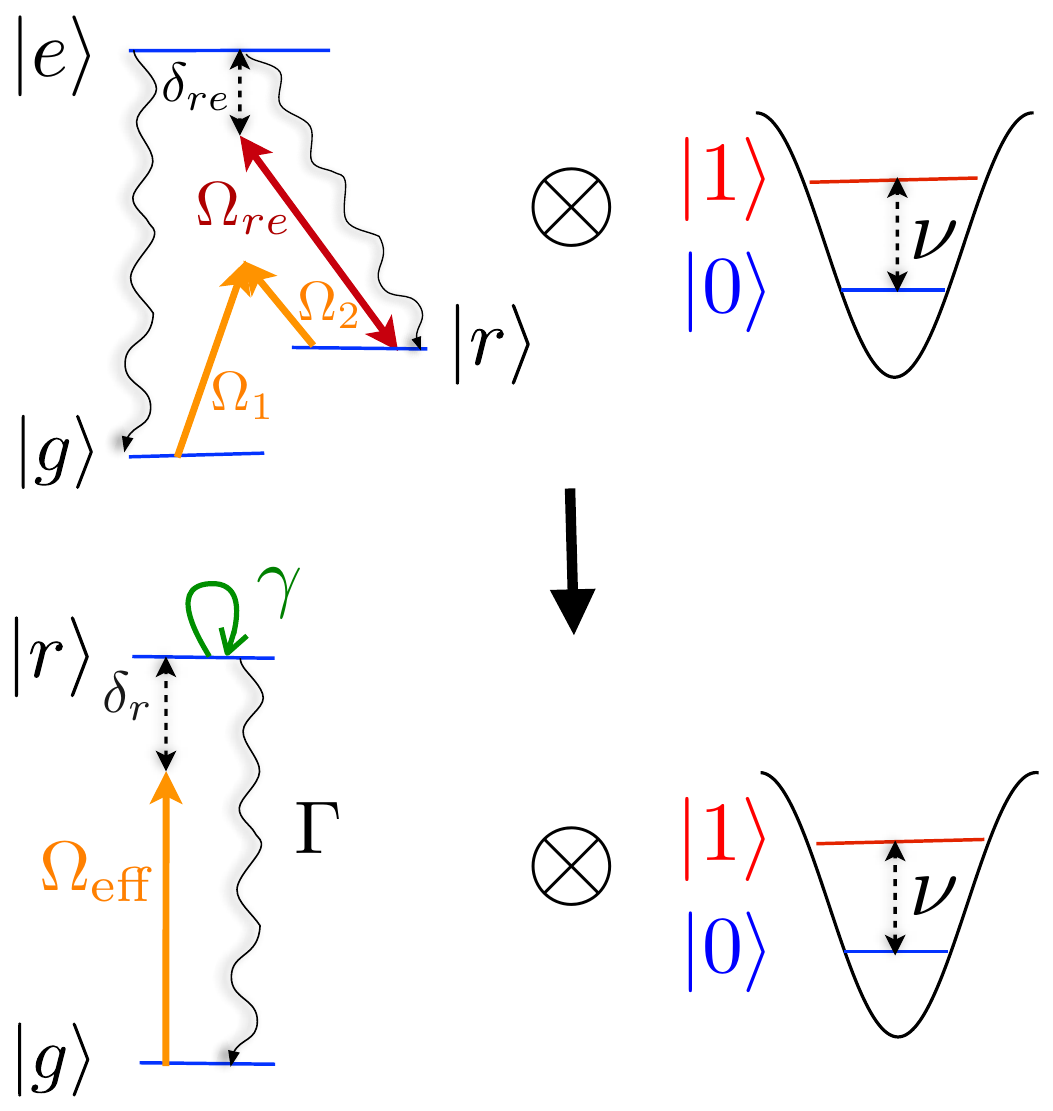}\hfill
 \caption{(color online) Relevant level structure and coupling/decay terms of single atom
   trapped at a lattice site. Upper part (Left): Internal levels of the atom:
   $\Lambda$ system $\ket{g}$, $\ket{r}$, $\ket{e}$, off-resonantly driven by
   lasers. (Right): Motional states in the lattice potential.  Lower part:
   After adiabatic elimination of $\ket{e}$, an effective two-level system
   with tunable decay rate $\Gamma$ and dephasing rate $\gamma$ is
   obtained.} \label{effectivetwolevel}
\end{figure}

The system we consider is an optical lattice populated with a single atomic
bosonic species.  We assume to be in the Mott-insulator regime with filling
factor one, where the on-site interaction is much larger than the tunneling
(hopping) between neighboring lattice sites. In this regime, the atoms are
localized such that each lattice potential is occupied with one atom. We aim
to use the motional ground and first excited state of the atom (denoted by
$\ket{0}, \ket{1}$\footnote{$\ket{0}_j$ denotes the localized Wannier function at site $j$, where $0$ is the band index}, respectively) to realize an effective spin-$1/2$ system in
each lattice site. To access the motional degree of freedom optically, we work
in the Lamb-Dicke regime where the motion of the atom is restricted to a
region small compared with the laser wavelength. We make use of the
anharmonicity of the lattice potential and, as explained in the following, of
decay of the atoms that leads to cooling of the system, to restrict the
dynamics to the two-dimensional subspace of $\{\ket{0}, \ket{1}\}$
\cite{Wi2002} (see Fig.~\ref{effectivetwolevel}). For the optical
manipulation, we assume that the atoms have internal degrees of freedom that
can be addressed with laser fields. We consider a $\Lambda$-scheme with two
ground states $\ket{g}$ and $\ket{r}$ (both trapped by the same optical
lattice potential) and an excited state $\ket{e}$. The level scheme of the
internal states of the atoms is shown in
Fig.~\ref{effectivetwolevel}. Off-resonant laser fields drive transitions
between the two ground states $\ket{g}$ and $\ket{r}$ and the excited state
$\ket{e}$. The system decays fast into the ground states, and, as we show
below, effectively decays into the state $\ket{g}$. Therefore, the atoms are
optically pumped to the state $\ket{g}\otimes\ket{0}$ and the states $\ket{r}$
and $\ket{e}$ can be adiabatically eliminated. Eliminating the excited state
$\ket{e}$ leads to the effective two-level system in the lower part of
Fig.~\ref{effectivetwolevel} with designable decay rates. Further elimination
of the state $\ket{r}$ leads to an effective description in the internal
ground state $\ket{g}$. The optical couplings by laser fields give rise to
effective Hamiltonians and effective dissipation (cooling) in the ground state
$\ket{g}$ at each lattice site. Details are given in Section
\ref{Diss}. In summary, we obtain an effective two-level system at each
lattice site with Hilbert space spanned by $\ket{g}\otimes\ket{0}$ and
$\ket{g}\otimes\ket{1}$ as depicted in Fig.~\ref{twolevelatom}. 

In the following sections, we show that engineering the optical couplings as
above leads to an effective master equation for the two-level system
$\ket{0}$, $\ket{1 }$ that describes (i) decay from $\ket{1}$ to $\ket{0}$ and
(ii) an effective magnetic field in $x$-direction. In the Mott insulator
regime, tunnel couplings between neighboring lattice wells can be treated as a
perturbation, which [iii] leads to an effective spin Hamiltonian. The
resulting master equation \footnote{for details see Eqn.~(\ref{eqndiss2})} is
given by
\begin{eqnarray}\label{eqndissmain}
\dot{\rho}_t=&\sum_k A^-(2\sigma_k^-\rho_t
\sigma_k^+-\{\sigma_k^+\sigma_k^-,\rho_t\}_+)-i\left[H
,\rho_t\right].
\end{eqnarray}
Here, $\sigma_k^+=\ketbra{1}{0}_k$ is the operator that excites an
atom at lattice site $k$ from the motional state $\ket{0}$ to state $\ket{1}$
and $\sigma_k^-=(\sigma_k^+)^\dag$.
The sum runs over all $N$ sites of the optical lattice potential.
The first part in Eqn.~(\ref{eqndissmain}) describes decay
from state $\ket{1}$ into state $\ket{0}$ as depicted in
Fig.~\ref{twolevelatom}. It is derived in Section \ref{Diss}. The decay parameter $A^-$ can be tuned by
changing the Rabi frequencies of the lasers and the detunings and is
given by Eqn.~(\ref{Apm2}) in Section \ref{All}. The Hamiltonian is
given by $H=H_B+H_{\textrm{spin}}$, where $H_B$ describes the
magnetic field in $x$-direction given by
\begin{eqnarray}\label{magneticfield}
H_B=\sum_k B_x(\sigma_k^++\sigma_k^-),
\end{eqnarray}
where $B_x$ is proportional to an effective magnetic field in $x$-direction. It is derived in
Section \ref{magnetic}. The Hamiltonian $H_{\textrm{spin}}$
describes the spin Hamiltonian
\begin{eqnarray}\label{Hspin}
H_{\textrm{spin}}=\sum_k\alpha_1(\sigma_k^x\sigma_{k+1}^x+\sigma_k^y\sigma_{k+1}^y)+\alpha_2
\sigma_k^z\sigma_{k+1}^z,
\end{eqnarray}
as derived in Section \ref{spin} from a tunnel coupling between neighboring
sites. The parameters $\alpha_1$ and $\alpha_2$ depend on the properties of
the optical lattice potential and can be tuned. Therefore, the Hamiltonian
$H_{\textrm{spin}}$ describes the XXZ model, the Ising model or the Heisenberg
model.
\begin{figure}
 \includegraphics[scale=0.6]{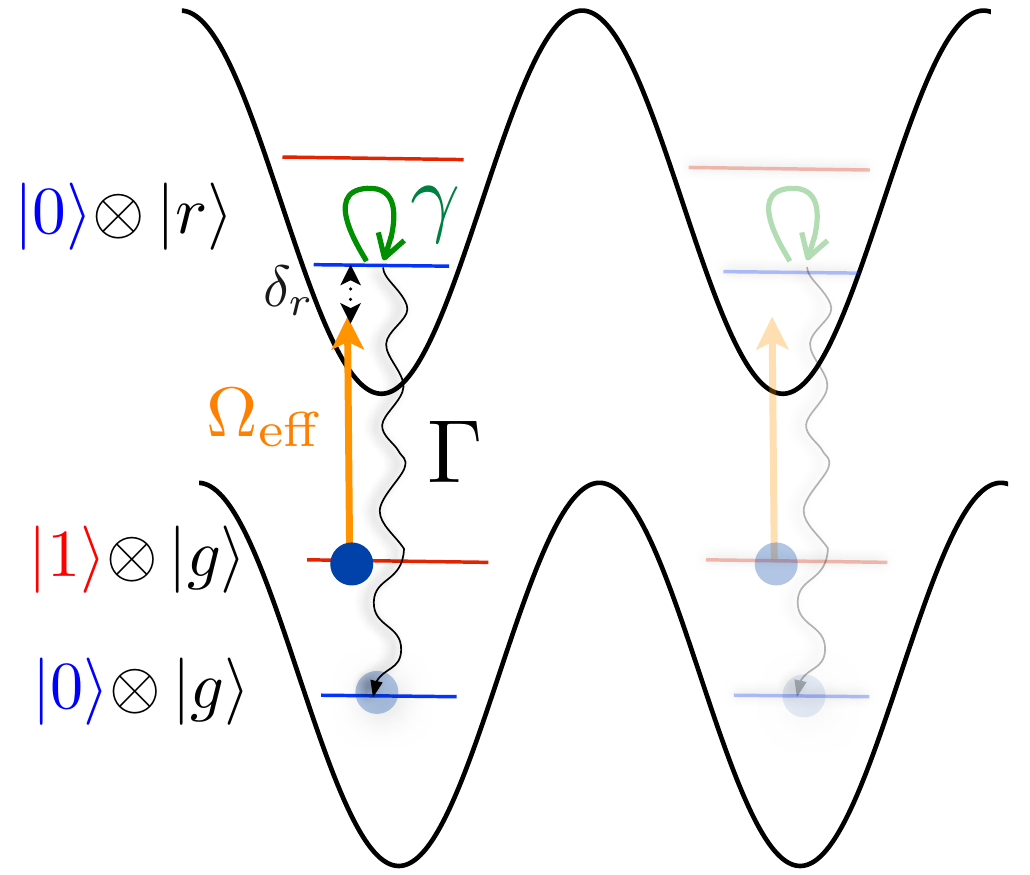}\hfill
\caption{(color online) Effective two-level system $\ket{g}$-$\ket{r}$ in the
optical lattice potential with motional states $\ket{0}$ and
$\ket{1}$. Choosing resonance conditions as explained in Section
\ref{Diss}, the atoms are selectively excited from
$\ket{g}\otimes\ket{1}$ to the state $\ket{r}\otimes\ket{0}$ and
spontaneously decay into $\ket{g}\otimes\ket{0}$.}
\label{decaytwolevel}
\end{figure}
In the following three Sections, we employ a perturbative approach
to derive a master equation comprising dissipation of Lindblad form
(i) as in Eqn.~(\ref{eqndissmain}), a magnetic field in
$x$-direction (ii) as in Eqn.~(\ref{magneticfield}) and an effective
spin Hamiltonian (iii) as in Eqn.~(\ref{Hspin}). For the sake of
clarity, we derive (i)-(iii) in three separate steps employing the
approximation of independent rates of
variation as explained in \cite{CDG92}.

\subsection{Optical couplings of internal atomic states: dissipation of Lindblad
form}\label{Diss}

In this Section, we show that optically addressing the atoms with
suitably tuned lasers allows to engineer decay as in Eqn.~(\ref{eqndissmain}).

We consider the internal levels $\ket{g}$, $\ket{r}$, $\ket{e}$ of
an atom at site $k$. The ground states $\ket{g}$ and $\ket{r}$
\begin{figure}
 \includegraphics[scale=0.3]{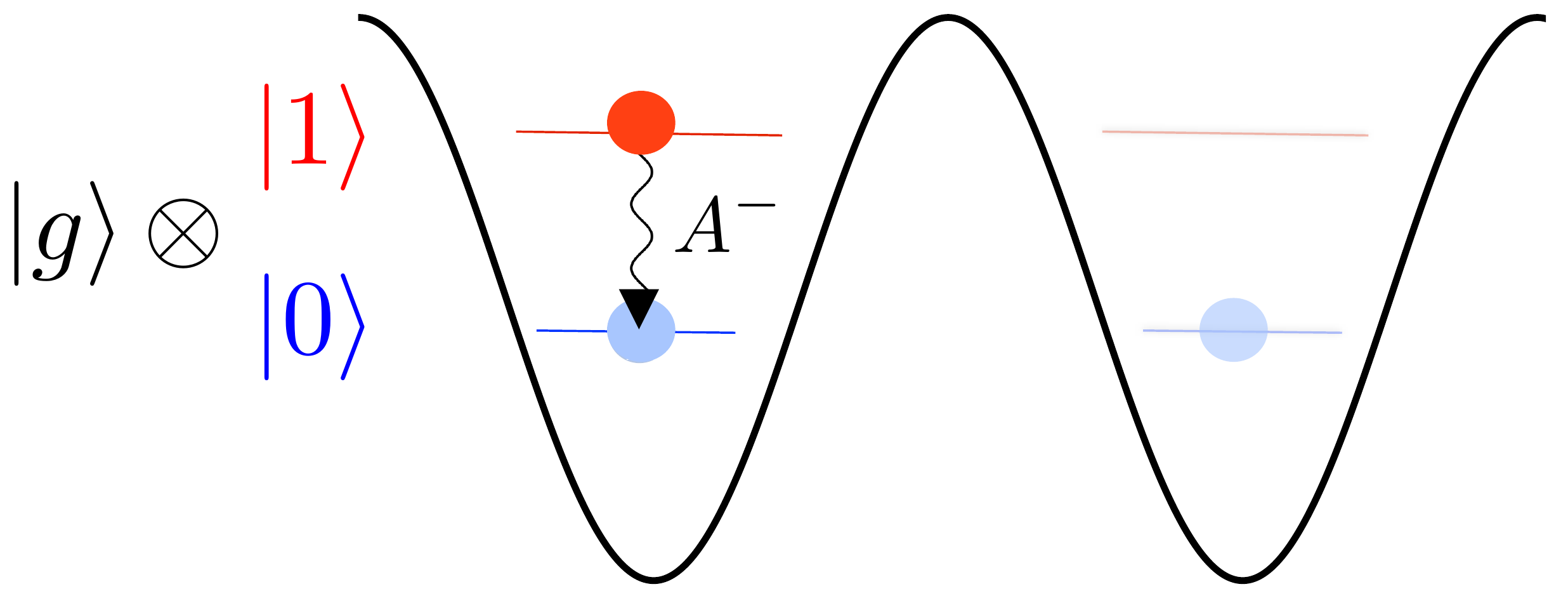}\hfill
\caption{(Color online) Decay of the effective two-level system $\ket{0}$,
$\ket{1}$ as described by the effective master equation derived in Section
\ref{Diss}. The dissipation strength $A^{-}$ is given in
Eqn.~(\ref{Apm}).} \label{twolevelatom}
\end{figure}
can be coupled via the excited state $\ket{e}$ by a detuned Raman
transition of two standing wave laser fields with Rabi frequencies
$\Omega_1$ and $\Omega_2$. Eliminating the excited state $\ket{e}$
leads to an effective coupling between $\ket{g}$ and $\ket{r}$ (see
Fig.~\ref{effectivetwolevel}) with
$\Omega_{\textrm{eff}}=\Omega_1\Omega_2/\delta_{re}$ where $\delta_{re}$ is
the detuning with respect to $\ket{e}$ (for details see Appendix
\ref{Apa}). To induce controlled dissipation, we couple $\ket{r}$ and
$\ket{e}$ by an additional off-resonant laser field (indicated by a red
arrow) in Fig.~\ref{effectivetwolevel}. Then adiabatic elimination of the
excited state $\ket{e}$ leads to an effective two-level system (as shown
in the lower part of Fig.~\ref{effectivetwolevel}) with states $\ket{r}$ and
$\ket{g}$ which has designable decay rates $\Gamma$ and $\gamma$ as derived in 
\cite{MaCi94} (see also Appendix \ref{Apa}). Thereby, the excited
state $\ket{e}$ that is broadened by spontaneous emission is
eliminated, and the effective two-level system $\ket{g}$-$\ket{r}$
allows to resolve the motional states $\ket{0}$ and $\ket{1}$ of the
lattice potential (note that we are in the Lamb-Dicke regime), as
can be seen in Fig.~\ref{decaytwolevel}. Under appropriate resonance
conditions that will be specified in the following, the atoms are
excited from state $\ket{1}\otimes\ket{g}$ to state
$\ket{0}\otimes\ket{r}$ and spontaneously decay into the state
$\ket{0}\otimes\ket{g}$ as shown in Fig.~\ref{decaytwolevel}.
Adiabatically eliminating the state $\ket{r}$, this corresponds to
an effective decay from state $\ket{1}\otimes\ket{g}$ into
$\ket{0}\otimes\ket{g}$. Thus the atoms effectively remain in the
internal ground state $\ket{g}$, such that the decay can be written
as an effective decay from state $\ket{1}$ to $\ket{0}$ as depicted
in Fig.~\ref{twolevelatom}.

In Appendix \ref{Apa}, we derive in a perturbative approach (that
corresponds to an adiabatic elimination of the state $\ket{r}$) a
master equation that describes the dynamics of the two-level system
$\ket{0}$, $\ket{1}$ of the atom. Assuming that the driving of level $\ket{r}$ is sufficiently weak such that
\begin{equation}
  \label{eq:cond1}
|\Omega_{\textrm{eff}}|\ll \Gamma, \gamma, \nu, |\delta_r|, 
\end{equation}
and
that the level broadening remains small 
\begin{equation}
  \label{eq:cond2}
  \Gamma+\gamma<\nu, 
\end{equation}
the master equation is given by
\begin{eqnarray}\label{Master}
\dot{\rho}_t=\sum_kA^-\left(2\sigma_k^-\rho_t
\sigma_k^+-\{\sigma_k^+\sigma_k^-,\rho_t\}_+\right)\notag\\+A^+\left(2\sigma_k^+\rho_t
\sigma_k^--\{\sigma_k^-\sigma_k^+,\rho_t\}_+\right)-i[H_{\textrm{eff}}^{(1)},\rho_t].
\end{eqnarray}  Here, $A^{+}$ determines the strength of
the heating terms and $A^{-}$ the strength of the decay terms.
For simplicity, $A^{\pm}$ are chosen to be independent of the
lattice site $k$. $A^{\pm}$ can be made dependent on the lattice
site $k$ by choosing different phases of the driving lasers as
explained in Appendix \ref{Apa}. Note that $A^+\ll A^-$ is required
for the validity of the approximation that restricts to the
$\ket{0}$ and $\ket{1}$ subspace.  $A^{-}$ and $A^{+}$ are
given by
\begin{equation}\label{Apm}
A^{\pm}=\Omega_{\textrm{eff}}^2\eta_{1}^2\frac{(\Gamma+\gamma)}{(\Gamma+\gamma)^2+(\delta_r\pm
\nu)^2}.
\end{equation} 
Here, $\delta_r$ is the effective detuning given by Eqn.~(\ref{deltar}) in
Appendix \ref{Apa}, $\eta_{1}=k_{1}/\sqrt{2M\nu}$ is the Lamb-Dicke parameter
where $k_1$ is the wave number of the laser with Rabi frequency $\Omega_1$,
$M$ the atomic mass, and $\nu$ denotes the energy difference between the
motional state $\ket{0}$ and $\ket{1}$ of the lattice potential. The
Hamiltonian $H_{\textrm{eff}}^{(1)}$ in the last term in Eqn.~(\ref{Master})
is given by
\begin{eqnarray}\label{Master2h}
H_{\textrm{eff}}^{(1)}=\sum_k\nu\ketbra{1}{1}_k+H_S,
\end{eqnarray} where $H_S$ describes AC Stark shifts on the motional levels that are
$\ll \nu$ and are given in more detail in Appendix \ref{Apa}. Now,
we have everything at hand to implement dissipation. If
$$\delta_r\approx \nu,$$ which can be achieved by choosing the laser
frequency $\omega_{l}$ in $\delta_r=\omega_r-\omega_{l}$
accordingly, the strength of the dissipation is much larger than the
strength of the heating:
\begin{equation}
  \label{eq:cond3}
  A^+\ll A^-. 
\end{equation}
Then, the master equation has only decaying terms and is of the form
\begin{eqnarray}\label{eqndiss}
\dot{\rho}_t=&\sum_k A^-(2\sigma_k^-\rho_t
\sigma_k^+-\{\sigma_k^+\sigma_k^-,\rho_t\}_+)-i\left[
H_{\textrm{eff}}^{(1)},\rho_t\right].
\end{eqnarray}
It describes decay of the atoms from state $\ket{1}$ into
$\ket{0}$, while the atoms effectively remain in the internal state
$\ket{g}$. By adiabatic elimination of the internal state $\ket{r}$,
we have thus shown that a master equation can be derived that can be
tuned such that it describes almost pure decay. 

\subsection{Optical couplings of internal atomic states: Effective magnetic field in
\texorpdfstring{$x$}{x}-direction}\label{magnetic}

To derive the effective magnetic field in $x$-direction, we consider
a detuned Raman transition. Two standing wave laser fields with Rabi
frequencies $\Omega_{a}$ and $\Omega_{b}$ couple the internal ground
state $\ket{g}$ and the excited state $\ket{e}$ of the atoms as
\begin{figure}
\includegraphics[scale=0.6]{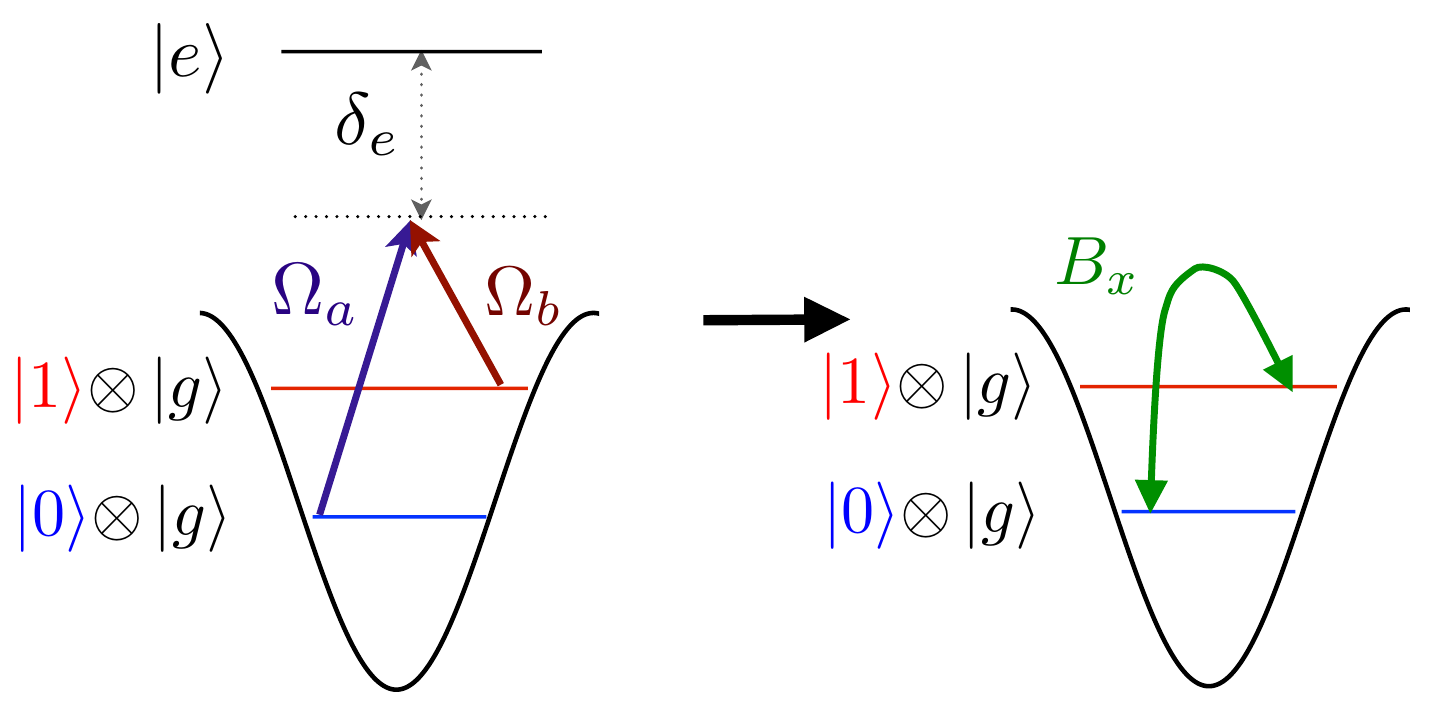}\hfill
\caption{(color online) Level scheme and transitions used to implement the transverse magnetic field. Left: A detuned Raman transition couples the internal
ground state $\ket{g}$ and the excited state $\ket{e}$ of the atom.
Right: Adiabatic elimination of the excited state $\ket{e}$ leads to
an effective magnetic field in $x$-direction (see Section
\ref{magnetic}) which drives transitions between the motional states
$\ket{0}$ and $\ket{1}$.}\label{Bfeldx}
\end{figure}
depicted in Fig.~\ref{Bfeldx}. The coupling is described by the
Hamiltonian
\begin{eqnarray}\label{Laser}
H_{ab}=&\sum_k\Omega_{a}\cos{(k_{a}x_k)}\ketbra{e}{g}_k\notag\\&+\Omega_{b}\sin{(k_{b}x_k)}\ketbra{e}{g}_k+\textrm{h.c.},
\end{eqnarray}
where $k_{a}$, $k_{b}$ denote the wave numbers of the lasers and
$x_k$ the displacement from the equilibrium position of the atom at
lattice site $k$. As we are in the Lamb-Dicke regime,
$\sin{(k_{b}x_k)}\approx\eta_{b}(\sigma_k^-+\sigma_k^+)$ \footnote{Note, that $\sin{(k_{b}x_k)}\approx\eta_{b}(c_k+c_k^{\dagger})$ where $c_k$ are bosonic operators that describe the harmonic oscillator states of the trapping potential. As explained before, we work in the truncated subspace of $\ket{0}$ and $\ket{1}$ due to the anharmonicity of the trap and the cooling to the ground state such that the $\eta_{b}(c_k+c_k^{\dagger})=\eta_{b}(\sigma_k^-+\sigma_k^+)$} and
$\cos{(k_{a}x_k)}\approx 1$. Under the condition
\begin{equation}
  \label{eq:cond4}
  \left|\Omega_{a}\right|,\left|\Omega_{b}\right|\ll|\delta_e|,
\end{equation}
where $\delta_e$ is the detuning of the driving lasers as depicted in Fig.~\ref{Bfeldx}, the excited state $\ket{e}$ can be adiabatically eliminated and we
get an effective Hamiltonian
\begin{equation}\label{Heff2}
H_{\textrm{eff}}^{(2)}=H_B=\sum_k B_x\sigma_k^x,
\end{equation}
that describes a tunable magnetic field in $x$-direction, where
$B_x$ is proportional to the effective magnetic field strength in $x$-direction is given by
$$B_x=\frac{2\Omega_{a}\Omega_{b}\eta_b}{\delta_e}.$$ 

Thus, we have derived an effective magnetic field in $x$-direction
that drives transitions between the motional states $\ket{0}$ and
$\ket{1}$ (as depicted in Fig.~\ref{Bfeldx} on the right), while the
atoms remain in the internal ground state $\ket{g}$.
\begin{figure}
 \includegraphics[scale=0.6]{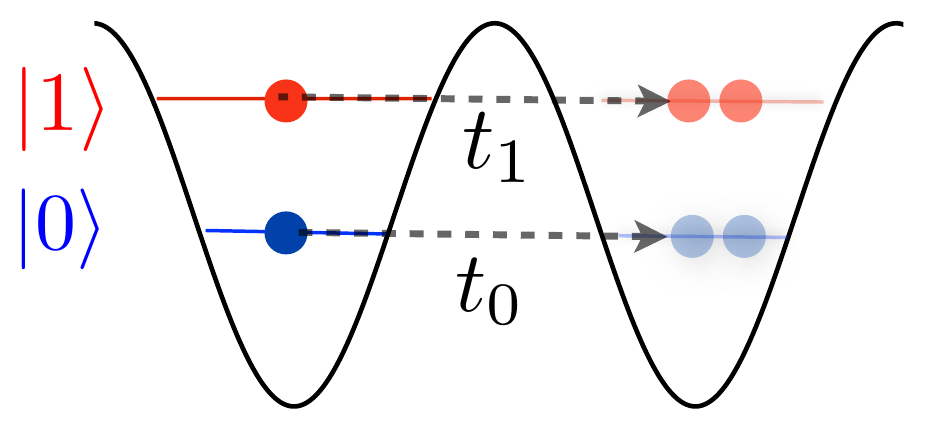}\hfill
  \caption{(color online) Tunneling between
neighboring lattice wells with tunnel amplitudes $t_0$ and $t_1$.
States with two atoms per lattice well are treated in perturbation
theory in Section \ref{spin}, as the on-site interaction is much
larger than the tunneling amplitudes.}\label{tunnel}
\end{figure}

\subsection{Effective spin Hamiltonian}\label{spin}
In the Mott-Insulator regime, bosonic atoms trapped by a lattice
potential with two motional states are described by the two-band
Bose-Hubbard model \cite{JBC+98} (see Appendix \ref{Apb}). We denote the on-site
interaction by $U_{01}$, $U_{00}$ and $U_{11}$ \footnote{$U_{xx'}$ is
the on-site repulsion of two atoms on lattice site $k$, where one
atom is in motional state $\ket{x}$ and the other one in $\ket{x'}$
with $x,x'=0,1$, respectively} and by $t_0$ ($t_1$) the amplitudes
for atoms in state $\ket{0}$ ($\ket{1}$) to tunnel to neighboring
lattice sites. We assume that the on-site interaction $U_{01}$,
$U_{00}$, $U_{11}$ $\gg$ $t_0$, $t_1$ such that tunneling between
neighboring wells that leads to states with two atoms in one lattice
well can be treated as a perturbation (see Fig.~\ref{tunnel}). Using
second order perturbation theory \cite{CDG92} (for a detailed
derivation see Appendix \ref{Apb}), we derive an effective spin
Hamiltonian $H_{\textrm{spin}}$ given by:
\begin{eqnarray}\label{Heff3}
H_{\textrm{eff}}^{(3)}=H_{\textrm{spin}}+B_z\sum_k\ketbra{1}{1}_k,
\end{eqnarray}
with
\begin{eqnarray}\label{eqn11bc}
H_{\textrm{spin}}=\sum_{k}\alpha_1(\sigma_k^x\sigma_{k+1}^x +
\sigma_k^y\sigma_{k+1}^y)+\alpha_2  
\sigma_k^z\sigma_{k+1}^z,
\end{eqnarray}
where $\alpha_1=-4t_0t_1/U_{01}$,
$\alpha_2=2[(t_0^2+t_1^2)/(2U_{01})-t_0^2/U_{00}-t_1^2/U_{11}],$ the magnetic
field in $z$-direction
\begin{equation}
  \label{eq:1}
  B_z=t_0^2/U_{00} -t_1^2/U_{11},
\end{equation}
using the Pauli spin matrices $\sigma_k^x$, $\sigma_k^y$ with
$\sigma_k^x=(\ketbra{0}{1}_k+\ketbra{1}{0}_k)/2$.  The Hamiltonian given
by Eqn.~(\ref{Heff3}) is an effective spin Hamiltonian that is
tunable by changing the lattice properties. If $\alpha_1$,
$\alpha_2>0$, $H_{\textrm{eff}}^{(3)}$ corresponds to the XXZ model
with a magnetic field in $z$-direction. If the lattice properties
can be tuned such that one of the tunneling constants $t_0$ or
$t_1\rightarrow0$, $H_{\textrm{spin}}$ is an Ising Hamiltonian with
a magnetic field in $z$-direction. For $\alpha_1=\alpha_2$, $H_{\textrm{spin}}$ corresponds to
the Heisenberg model.
\begin{figure}
 \includegraphics[scale=0.75]{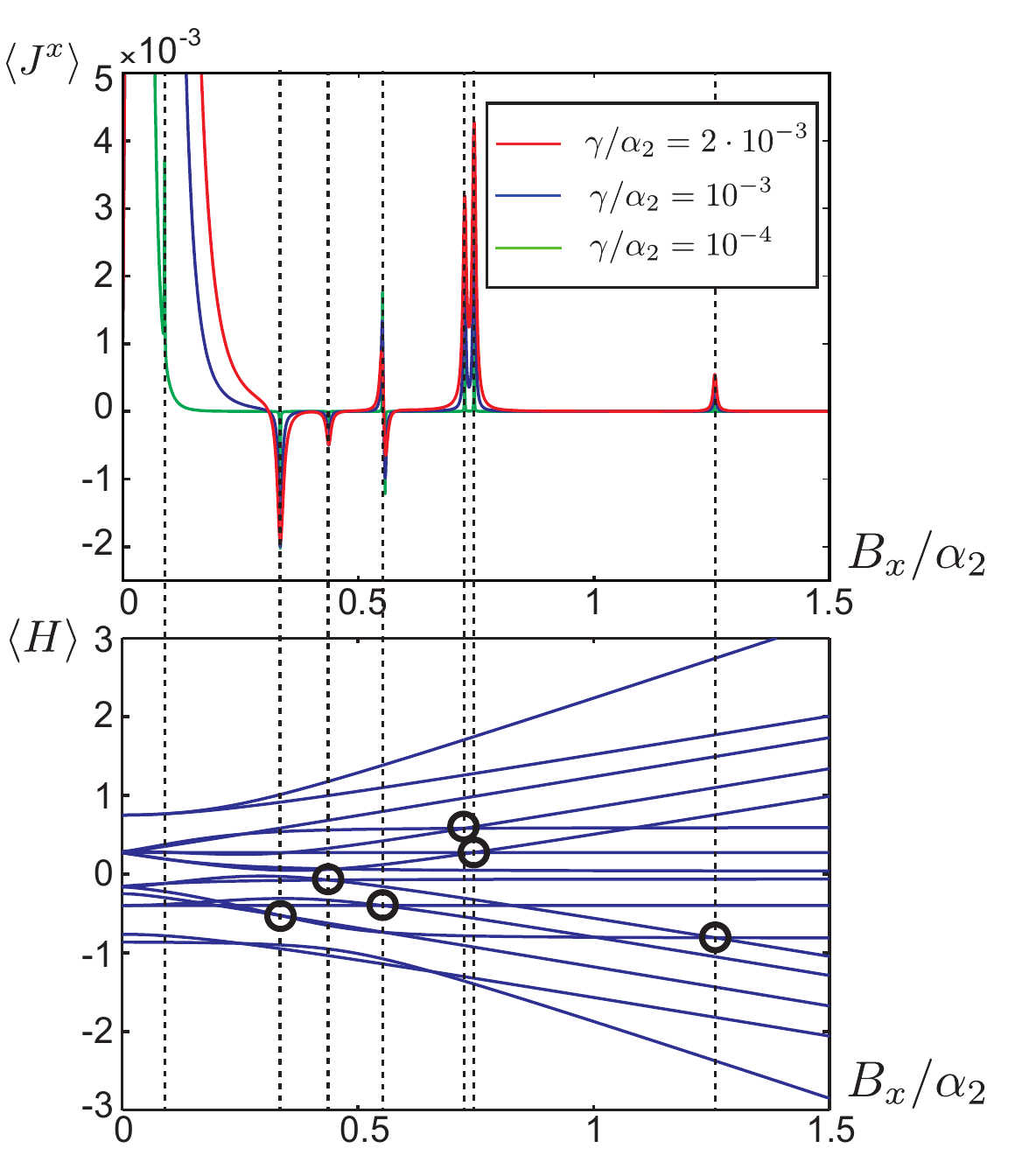}
 \caption{(color online) XXZ model with 4 spins, $\alpha_1=\frac{1}{4} \alpha_2$ and open
   boundary conditions under local dissipation of form given by
   Eqn.~(\ref{eqndiss2}). Upper part: Steady-state expectation value $\langle
   J^x\rangle$ plotted versus the magnetic field $B_x/\alpha_2$.  Peaks are
   observed that narrow for decreasing the dissipation strength. Lower part:
   Spectrum of the XXZ chain in the magnetic field $B_x$ plotted versus
   $B_x/\alpha_2$. Peaks in the steady-state expectation value (upper part)
   appear at crossing points of the Hamiltonian that are marked with black
   circles.  }\label{4spinsxxynichtgener}
\end{figure}
\subsection{Dissipative one-dimensional spin chain in a magnetic field}\label{All}
In the previous Sections, we showed --- for the sake of clarity in separate
steps --- that optical couplings of the internal levels can be engineered such
that we obtain a master equation of Lindblad form [Eqn.~(\ref{Master})]
fulfilling the demands (i)-(iii) of tunable dissipation, spin-interaction, and
tunable transverse field.  Combining these results, one has to carefully
consider the order of magnitude of each term. Doing so, we find that the
magnetic field in $z$-direction $B_z$ in Eqn.~(\ref{Heff3}) and the Stark
shifts in Eqn.~(\ref{Master2h}) can be of the same order of magnitude as
$\nu$. Stark shifts and $B_z$ lead to an effective energy difference between
the motional states $\ket{0}$ and $\ket{1}$ given by
$$\tilde{\nu}=\nu+B_z+s_--s_+,$$ where $B_z$ is defined in
Eqn.~(\ref{eqn11bc}) and $s_-$, $s_+$ are the AC Stark shifts in
Eqn.~(\ref{Master2h}) (see Appendix \ref{Apa}). Therefore, combining all
results, the laser detuning $\delta_r$ that enters in the $A^{\pm}$ has to be
adjusted to $\tilde{\delta}_r$ such that
$\tilde{\delta}_r-\nu=\delta_r-\tilde{\nu}$ which means that
$\tilde{\delta}_r=\delta_r\pm(B_z+s_--s_+)$.

Then, combining the results from Eqns.
(\ref{Master}), (\ref{Heff2}) and (\ref{Heff3}), the master equation reads
\begin{eqnarray}
&\dot{\rho}_t=\sum_kA^+(2\sigma_k^+\rho_t
\sigma_k^--\{\sigma_k^-\sigma_k^+,\rho_t\}_+)\notag\\&+A^-(2\sigma_k^-\rho_t
\sigma_k^+-\{\sigma_k^+\sigma_k^-,\rho_t\}_+)-i[H,\rho_t],\label{eq:finalMEq}
\end{eqnarray}
where the rates $A^{\pm}$ are modified by the renormalized $\tilde{\delta}_r$:
\begin{equation}\label{Apm2}
A^{\pm}=\Omega_{\textrm{eff}}^2\eta_{1}^2\frac{(\Gamma+\gamma)}{(\Gamma+\gamma)^2+(\tilde{\delta}_r\pm
\nu)^2}.
\end{equation}
The Hamiltonian part of the master equation is given by
\begin{eqnarray}
H=&H_{\textrm{spin}}+H_B+\tilde{\nu}\sum_k\ketbra{1}{1}_k,
\end{eqnarray}
where $H_{\textrm{spin}}$ is given by Eqn.~(\ref{eqn11bc}) and $H_B$ by
Eqn.~(\ref{Heff2}). The magnetic field in $z$-direction and Stark shifts have
been included in $\tilde{\nu}$. For $\tilde{\delta}_r\approx \nu$, as shown
before, decay dominates over heating: $A^-\gg A^+$. Then, the master equation
has only decaying terms and Eqn.~(\ref{eq:finalMEq}) describes a dissipative
XXZ spin chain in a magnetic field with both $x$ and $z$ components. However,
only $B_x$ is fully tunable, while $B_z$ is large (compared to $B_x, A^\pm$)
and required to be so by the conditions for adiabatic elimination,
cf. Eqn.~(\ref{eq:cond1}). However, an effective dissipative XXZ chain without
any field in $z$-direction would be advantageous for observing critical
behavior in the steady-state dynamics that we study in the next Sections.
Therefore, we transform to a frame rotating with $\tilde{\nu}$. In the
rotating frame, $H_B$ becomes time-dependent. To obtain a time-independent
field in $x$-direction, the detuned Raman lasers that lead to the effective
magnetic field $B_x$ have to be chosen time-dependent, adapted to the
rotating frame (i.e., suitably detuned from the two-photon resonance). This
then yields a time-independent transversal magnetic field, and the master
equation in the rotating frame is then given by
\begin{eqnarray}\label{eqndiss2}
\dot{\rho}_t=&\sum_k A^-(2\sigma_k^-\rho_t
\sigma_k^+-\{\sigma_k^+\sigma_k^-,\rho_t\}_+)\notag\\&-i\left[
H_{\textrm{spin}}+H_B,\rho_t\right].
\end{eqnarray}
It corresponds to the master equation given by Eqn.~(\ref{eqndissmain}).  In
summary we have shown how to implement a one-dimensional spin chain with
nearest-neighbor interaction described by the XXZ or the Ising model and a
tunable effective magnetic field in $x$-direction under dissipation. This
system is an ideal testbed for studying steady-state dynamics of dissipative
spin models as discussed in the next Section. Note, that since we are in a
rotating frame, observables other than the collective spin operator $\langle
J_z\rangle $ become explicitly time-dependent.

\subsection{Steady-state behavior: Discontinuous steady-state behavior related to spectrum of Hamiltonian}
A particular important characterization of dissipative dynamics is through
their steady state: if it is unique (or distinguished by some conserved
quantity) it allows for robust preparation of these states. Abrupt changes in
the steady state as system parameters are varied may signal dissipative
quantum phase transitions
\cite{Dimer2007,MoPa07,Diehl2008,EiPr11,Werner2005,Mueller2012,Znidaric11}.
We study the steady-state behavior of short spin chains under dissipation in a
magnetic field in $x$-direction with numerical simulations. We find that the
one-dimensional XXZ model with 4 spins as given by Eqn.~(\ref{eqn11bc}), where
we chose as a typical example $\alpha_1=\frac{1}{4}\alpha_2$, shows a surprising
behavior: Changing the external magnetic field in $x$-direction, peaks occur
in the steady-state expectation values of the collective spin operators
$J^{x/z}=\sum_k \sigma_k^{x/z}$ for weak dissipation, see
Fig.~\ref{4spinsxxynichtgener}. Here, we have considered dissipation as in
Eqn.~(\ref{eqndiss2}) with equal dissipation strength on each spin. We find
that decreasing the strength of the dissipation the peaks become more
narrow and each peaks height approaches a finite value. For small $\gamma$
we observe very narrow peaks. This indicates a discontinuity in the
steady-state expectation values of the spin operators. We find, that these
narrow peaks appear exactly at points where the
Hamiltonian becomes degenerate. 
In the following Section we study this phenomenon in more
generality.
\begin{figure}[t]
\includegraphics[scale=0.75]{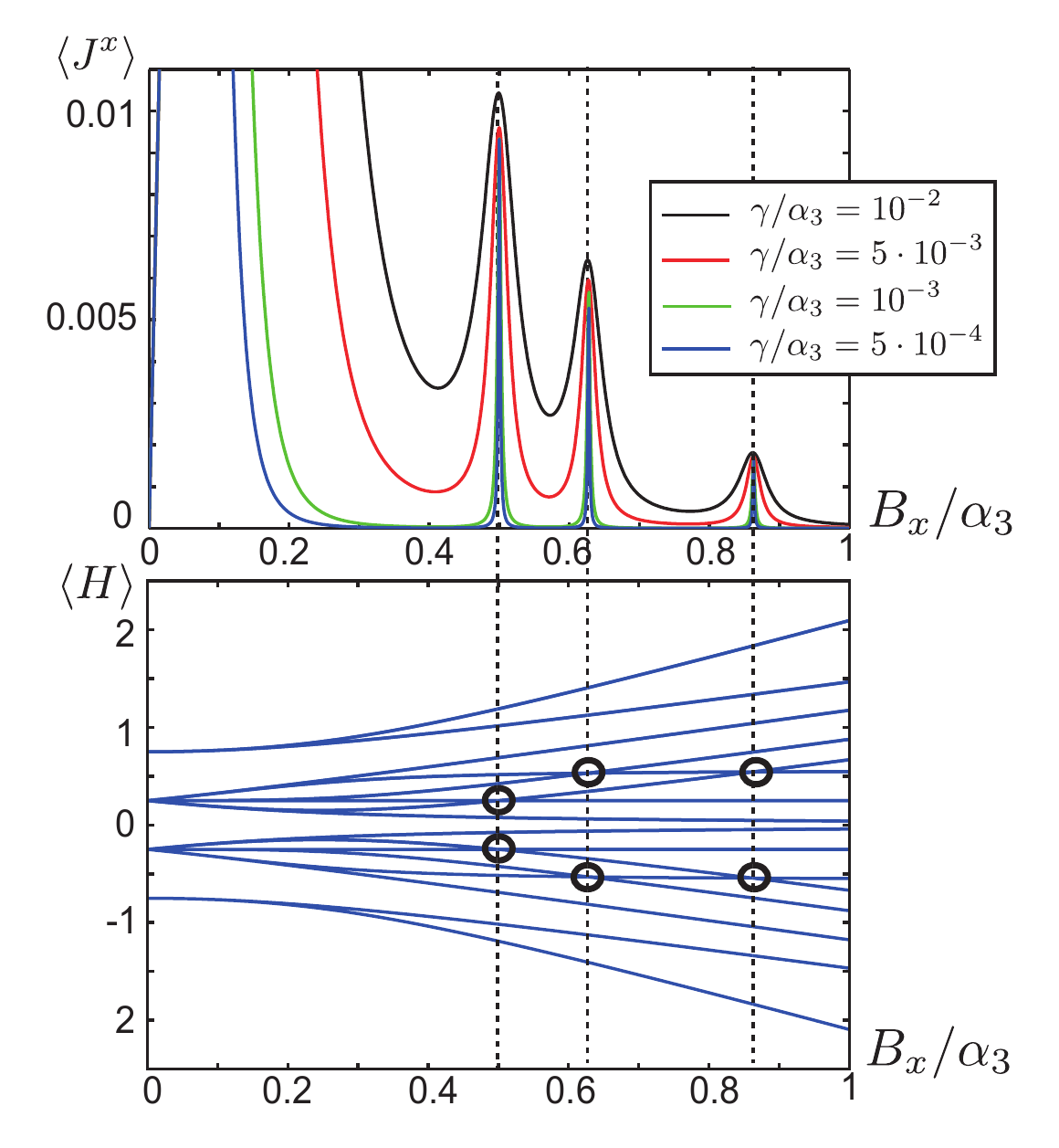}
\caption{(Color online) 
Ising model with 4 spins with open boundary conditions in a transverse
magnetic field $B_x$ and under local dissipation of form given by
Eq.~(\ref{eqn3}. Upper part: Steady-state expectation value $\langle J^x\rangle$
plotted versus $B_x/\alpha_3$.  Peaks are observed that become more narrow for
decreasing dissipation strength. Lower part: Spectrum of the Hamiltonian
plotted versus $B_x/\alpha_3$. Peaks in the steady-state expectation value
(upper part) appear at degeneracy points of the Hamiltonian that are marked
with black circles.
}
  \label{4spinsising}
\end{figure}
\begin{figure}[t]
\includegraphics[scale=0.75]{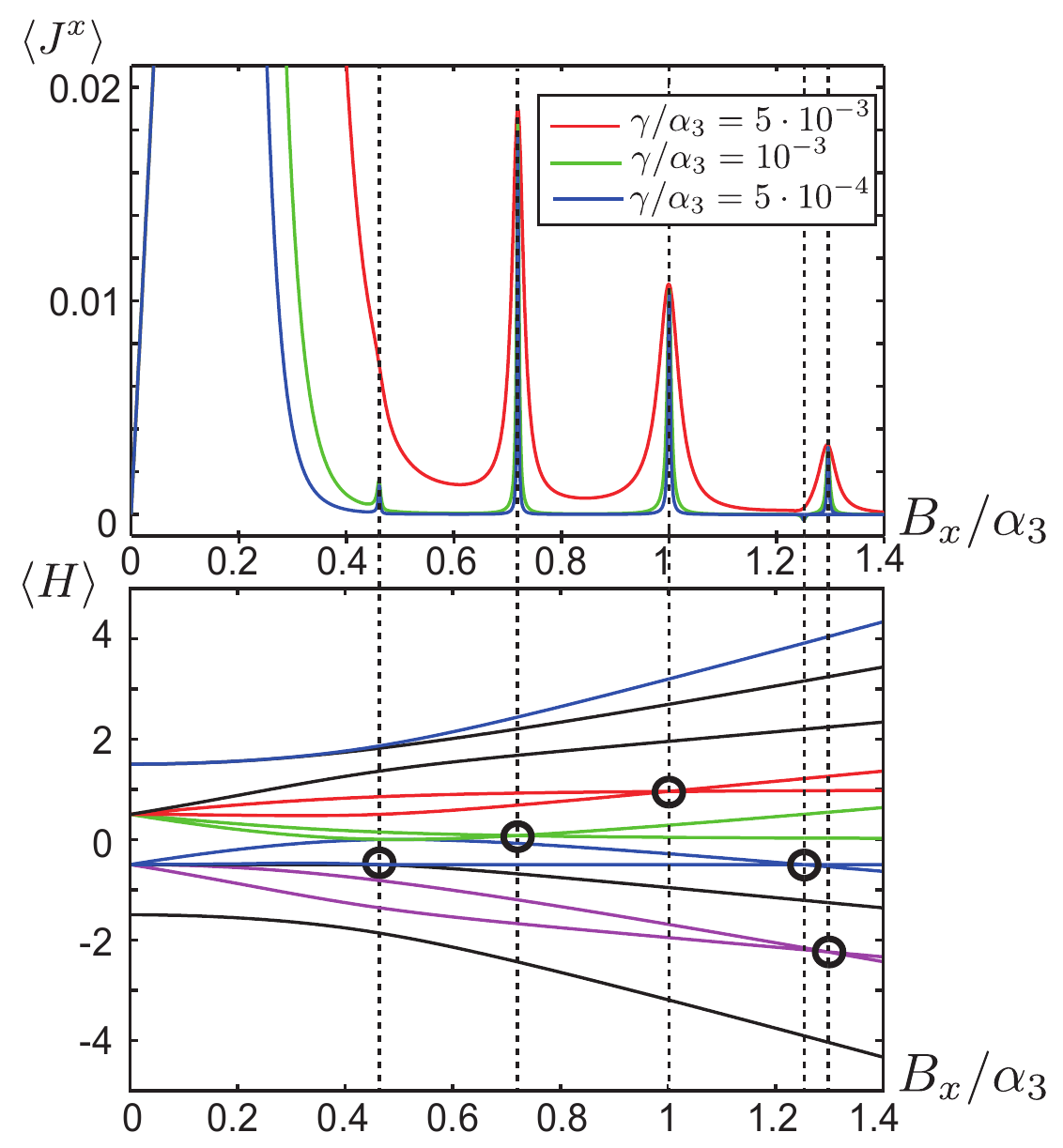}
\caption{(Color online) Ising model with 6 spins with periodic boundary
conditions in a magnetic field $B_x$ and under collective dissipation
of the form given by Eq.~(\ref{eqncoll}) in the translation and reflection
symmetric subspace $T=R=1$. Upper part: Steady-state expectation value
$\langle J^x\rangle$ plotted versus $B_x/\alpha_3$.  Lower part:
Spectrum of the Hamiltonian plotted versus $B_x/\alpha_3$.
}
  \label{6spinsising}
\end{figure}

\section{Discontinuities in the steady-state dynamics of a general class of
  one-dimensional spin models under dissipation}
In the previous Section we saw that for the one-dimensional XXZ
model, peaks in the steady-state expectation values of the collective spin
operators appear, that are closely related to the spectrum of the
Hamiltonian. In the following, we study in more generality,
independent of a physical implementation, local one-dimensional spin
Hamiltonians under dissipation of different kinds. We present a
condition that elucidates the discontinuous behavior of the steady
state at degeneracy points of the Hamiltonian. Then, we study and explain this
condition in more detail for Ising Hamiltonians.

\subsection{Numerical studies of discontinuous behavior in the steady
state}

We numerically simulate short spin chains. First, we study the one-dimensional
Ising model with open boundary conditions, described by the Hamiltonian
\begin{equation}\label{eqn1aneu}
H=H_{zz}+H_B
\end{equation}
with
\begin{equation}\label{eqn1a}
H_{zz}=\alpha_3\sum_k \sigma_k^z\sigma_{k+1}^z,
\end{equation}
and $H_B$ as in Eqn.~(\ref{magneticfield})
subject to local or collective decay with Lindblad operators $\propto
\sigma_k^-$ or $\sum_k\sigma_k^-$, respectively. The master equation
describing the full system with local dissipation is given by
\begin{equation}\label{eqn3}
\dot{\rho}_t=\sum_k \gamma_k \left(2\sigma_k^-\rho_t
\sigma_k^+-\{\sigma_k^+\sigma_k^-,\rho_t\}_+\right)-i[H,\rho_t].
\end{equation}

Changing the magnetic field $B_x$, we find that for weak dissipation the
steady-state expectation values of the spin operators $\langle J^x\rangle$ and
$\langle J^z\rangle$ change abruptly at particular values of $B_x$, see
Fig.~\ref{4spinsising}. Here, we have considered dissipation as in
Eqn.~(\ref{eqn3}) with equal dissipation strength on each spin,
$\gamma_k=\gamma$. Decreasing the strength of the dissipation, i.e.,
decreasing $\gamma$, the peaks become more narrow and their height converges
to some finite value, while the expectation value vanishes elsewhere. For
$\gamma\rightarrow 0$, we observe very narrow peaks, which indicate
discontinuities in the steady-state expectation values of the spin operators.
We find, that these narrow peaks appear only at degeneracy
points of the spectrum of the Hamiltonian. I.e., to every peak found at some
value of $B_x=x_0$ for $\gamma\rightarrow 0$, at least one pair of
degenerate
eigenvalues $\lambda_1, \lambda_2$ of the local spin Hamiltonian $H_{zz}$ can
be found, i.e., $\lambda_1(x)=\lambda_2(x)$ at $x=x_0$.  Note that the
discontinuities in the steady state at critical system parameters are only
observed for $\gamma\neq 0$. I.e., the (weak) dissipation allows us to gain
information about the Hamiltonian's properties that is not readily accessible
in the case of $\gamma=0$.

This effect can be observed for
different kinds of spin Hamiltonians such as, for example, the XXZ model (see
Fig.~\ref{4spinsxxynichtgener}), both for periodic and open boundary
conditions. Moreover, changing the type of dissipation, the observed behavior
does not change qualitatively. E.g., collective dissipation, which describes
the dynamics of spins all coupled to the same bath and leads to the
master equation
\begin{equation}\label{eqncoll}
\dot{\rho}_t=\gamma (2J^-\rho_t J^+-\{J^+J^-,\rho_t\}_+)-i[H,\rho_t],
\end{equation}
where $J^{\pm}=\sum_k \sigma_k^{\pm}$, also
leads to discontinuous behavior in the steady-state expectation
values as shown in Fig.~\ref{6spinsising} for the Ising model. Choosing an
"inhomogeneous" dissipation which is of the form of the dissipative part
in Eqn.~(\ref{eqn3}), where now the strengths of the dissipation
$\gamma_k$ are different for each spin, peaks can be observed for an
even larger class of spin Hamiltonians: For $\gamma_k=\gamma$, and
$H=H_H+H_B$, where $H_H$ is the Heisenberg spin Hamiltonian, we do
not observe any peaks. However, if we choose different dissipation
strengths $\gamma_k$ for each spin, we find peaks at the degeneracy
points of the Hamiltonian, as can be seen in
Fig.~\ref{openheisenberg}.

\subsection{General condition for discontinuities in the steady state}

Since the Liouvillian depends smoothly on the system parameters, the
observed discontinuities must be related to degeneracies in the
spectrum of $\cal L$. As we shall see, in the weak dissipation limit
they are directly related to degeneracy points of the Hamiltonian.

We consider a system described by the master equation
\begin{equation}\label{eqn41}
  \dot{\rho}(t)={\cal L}\rho \equiv
  \left[\mathcal{L}_0(x)+\gamma\mathcal{L}_1\right]\rho(t),
\end{equation}
where $$\mathcal{L}_0(x)(\rho)=-i[H(x),\rho],$$ with a Hamiltonian $H(x)$
depending (analytically) on a parameter $x$. For simplicity, we consider the
case that $H_0(x)$ is non-degenerate for $x\not=x_0$. The term $\mathcal{L}_1$
contains dissipative terms and is independent of $x$. We are interested in the
limit of weak dissipation $\gamma\to0$ and in the change of the steady state
at the degeneracy point $x=x_0$.

The steady state $\rho_\mathrm{ss}(x)$ is determined by ${\cal
  L}(x)\rho_\mathrm{ss}(x)=0$ and can be determined perturbatively.  The
kernel of ${\cal L}_0(x)$ is highly degenerate, being spanned by all
eigenprojectors $\ketbra{\lambda_i(x)}{\lambda_i(x)}$ of the (non-degenerate)
$H_0(x)$. This degeneracy is lifted by ${\cal L}_1$ and the steady state for
$\gamma\to 0$ is for $x\not=x_0$ given by
\begin{equation}\label{eqn6a}
\mathrm{P}^D(x)\mathcal{L}_1\mathrm{P}^D(x)\rho_{ss}(x)=0,
\end{equation}
where
\begin{equation}\label{eqn6P}
\mathrm{P}^D(x)\rho =\sum_i\ketbra{\lambda_i(x)}{\lambda_i(x)}\rho
\ketbra{\lambda_i(x)}{\lambda_i(x)}.
\end{equation}

\begin{figure}
    \includegraphics[scale=0.75]{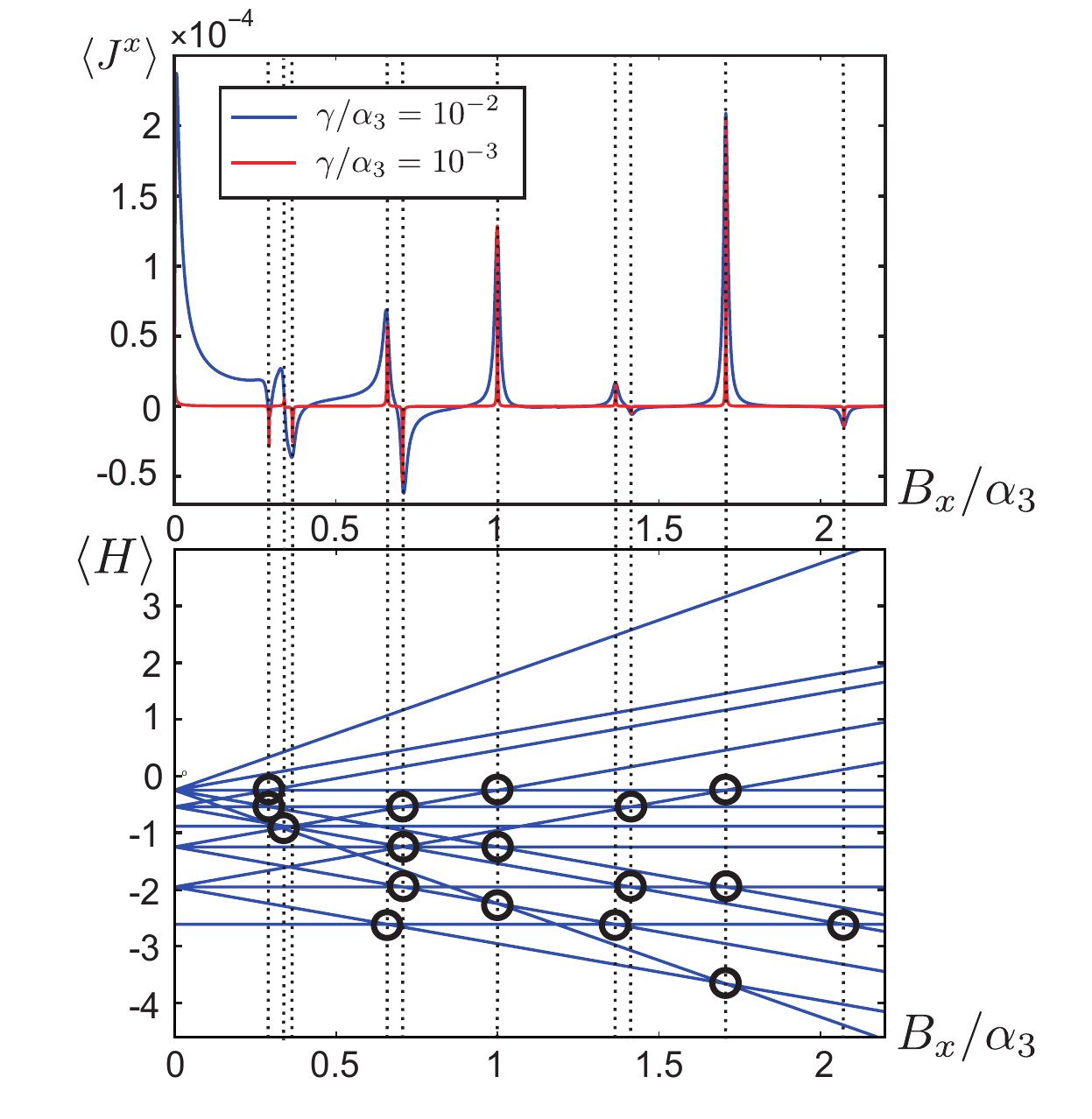}
  \caption{
(Color online) Heisenberg model with 4 spins with open boundary
conditions in a magnetic field $B_x$ and under local dissipation as
given by Eq.~(\ref{eqn3}) with different dissipation strengths
$\gamma_k$. Upper panel: steady-state expectation value $\langle J^x
\rangle$ plotted versus the $B_x/\alpha_3$. Lower part: Spectrum of
the Hamiltonian plotted versus $B_x/\alpha_3$.
}
  \label{openheisenberg}
\end{figure}
The possibility of discontinuous behavior of $\rho_\mathrm{ss}(x)$
at $x=x_0$ arises from the enlargement of the kernel of ${\cal
L}_0(x)$ at this point: if $\lambda_i$ and $\lambda_j$ become degenerate at $x=x_0$ then coherences between the corresponding 
eigenvectors (i.e., $\ketbra{\lambda_i(x))}{\lambda_j(x)}, i\not=j$) become
stationary at $x=x_0$. We denote by $P^{\Delta}$ the 
projector on these additional elements in the kernel of ${\cal
L}_0(x_0)$ \footnote{Here we use that if
  $H(x)$ is a holomorphic function of $x$ (we are typically concerned with
  linear dependence on $x$ only) the eigenvectors of $H(x)$ can be chosen as
  holomorphic (and thus continuous) functions of $x\in\mathbbm{R}$
  \cite{Kato}. Then $\lim_{x\to x_0}P^D(x_0)$ is well defined and we can
  define $P^{\Delta}$ as the difference of the projector on the kernel of
  ${\cal L}_0(x_0)$ and $\lim_{x\to x_0}P^D(x_0)$.}.
 As we show in Appendix \ref{Apc}, a discontinuity
$\rho_\mathrm{ss}(x_0)\not=\lim_{x\to
  x_0}\rho_\mathrm{ss}(x)$ arises if
\begin{equation}\label{conend1}
\mathrm{P}^{\Delta}\mathcal{L}_1\lim\limits_{x \rightarrow
x_0}\rho_{ss}(x)\neq0,
\end{equation}
i.e., if ${\cal L}_1$ couples the steady state to the newly available
subspace $P^{\Delta}$ in the kernel of ${\cal L}_0$. For simplicity, we made
the assumption that the Hamiltonian is non-degenerate for $x\neq x_0$. If the
Hamiltonian does have degeneracies outside $x_0$, but additional eigenvectors
become degenerate at $x=x_0$ the argumentation follows identical lines, as
also in this case, ${\cal L}_1$ can couple the steady state to a newly
available subspace $P^{\Delta}$.

\begin{figure}[h]
    \centering
\includegraphics[width=87mm]{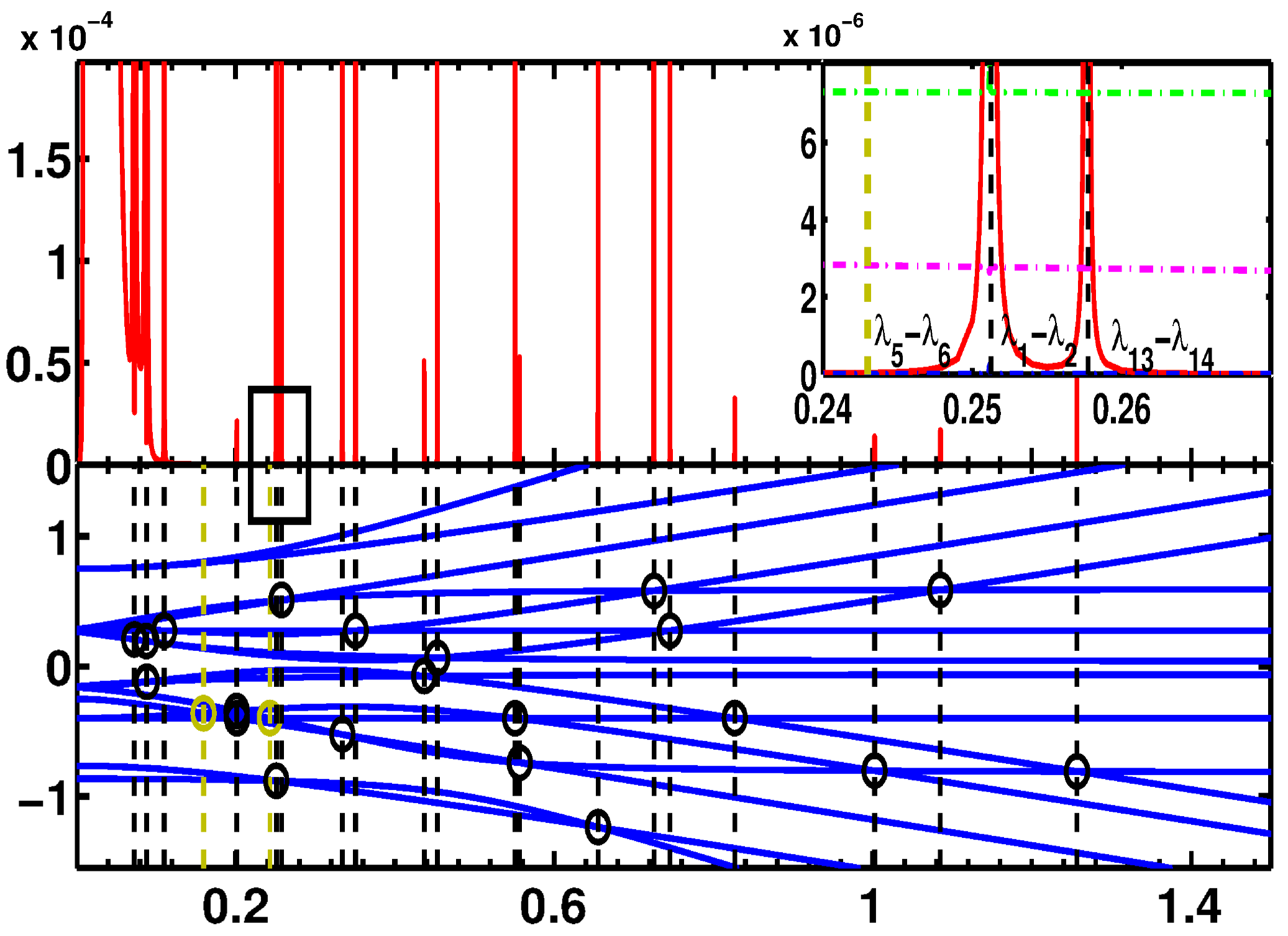}
\caption{(Upper part) Steady state infidelity $I_{\delta
      B}(B_x)$ (see text) for the 4-spin XXZ-model in transverse field $B_x$
    with local dissipation (cf. Fig.~\ref{4spinsxxynichtgener}) for $\delta
    B=3\cdot 10^{-6}$ and weak dissipation $\gamma=0.5\cdot10^{-4}$.  (Lower
    part) Spectrum of $H_\mathrm{spin}$, dashed vertical lines indicate
    degeneracy points (encircled). Peaks in $I_{\delta B}$ line up with
    degeneracy points of $H(B_x)$, except for two
    (colored green) for which condition (\ref{conend1}) does not hold. \\
    The inset shows the vicinity of $B_x\approx0.25$, where three crossings of
    eigenvalues occur: $(\lambda_5,\lambda_6), (\lambda_1,\lambda_2)$, and
    $(\lambda_{13},\lambda_{14})$. The dash-dotted horizontal lines show the
    2-norm (scaled by $7\cdot10^{-7}$) of the left-hand side of
    Eq.~(\ref{conend1})
    $C_{i,j}=\|\mathrm{P}^{\Delta_{i,j}}\mathcal{L}_1\lim\limits_{x'
      \rightarrow x}\rho_{ss}(x')\|_2$ for the three relevant projectors
    $\mathrm{P}^{\Delta_{i,j}}, (i,j)=(1,2)$ (green), $=(5,6)$ (blue), and
    $(=(13,14))$ (magenta): $C_{5,6}$ vanishes at the crossing of
    $(\lambda_5,\lambda_6)$, hence there is no peak in $I_{\delta B}$, while
    the other two lead to a peak in $I_{\delta B}$, since $C_{i,j}$ is
    finite.}
\label{fig:discont}
\end{figure}

Let us have another look at Figs.~\ref{4spinsxxynichtgener} -
  \ref{openheisenberg} in the light of the previous paragraph. Clearly, all
  the sharp isolated peaks occur for values of $B_x$ (which plays the role of
  the parameter $x$), at which a degeneracy occurs, satisfying a necessary
  condition for the Eqn.~(\ref{conend1}). However, not all degeneracy points
  lead to discernible peaks, e.g., in Fig.~\ref{4spinsxxynichtgener}. This can
  show that ${\cal L}_1$ does not couple the steady state to $P^\Delta$ or
  that the discontinuity is not be witnessed by the expectation value of
  $J^x$. For most peaks studied here, however, the reason is simply that the
  corresponding peaks are too small and sharp to be resolved in the
  plot.\\
  These points are illustrated in Fig.~\ref{fig:discont}, which shows that the
  steady state changes abruptly at all degeneracy points of $H$ for the 4-spin
  XXZ-model with local dissipation except for two such points
  (at $B_x\approx0.16, 0.24$), where $\mathcal{L}_1$ does not couple to the
  coherences. To measure how quickly $\rho_\mathrm{ss}$ changes with $B_x$ we
  use (in analogy to the ground state fidelity introduced in \cite{ZaPa06} for
  the study of quantum phase transitions) the ``steady-state infidelity''
  $I_{\delta B}(B_x)\equiv1-F(\rho(B_x),\rho(B_x+\delta B))$. Here
  $F(\rho,\sigma)=\mathrm{tr}(\sqrt{\sigma^{1/2}\rho\sigma^{1/2}})^2\in[0,1]$
  denotes the Uhlmann fidelity \cite{Uhl76} between two density matrices,
  which measures how similar $\rho$ and $\sigma$ are. Peaks in $I_{\delta
    B}(B_x)$ (for small $\delta B$) indicate that the steady state changes
  abruptly with $B_x$. For weak dissipation this happens close to all
  degeneracy points of the Hamiltonian when Eq.(\ref{conend1}) holds.

Note also, that in Figs.~\ref{4spinsxxynichtgener} - \ref{openheisenberg}, a
large feature appears in the steady-state expectation value $\langle J^x
\rangle$ around $B_x=0$. It narrows for decreasing $\gamma$, but is not a
sharp peak for any of the parameters used for $\gamma$. This broad peak
represents the effect of one (or several, cf.\ Figs.~\ref{4spinsxxynichtgener}
and \ref{openheisenberg}) unresolved degeneracies around $B_x=0$: Note that
for all spin models considered, the degeneracy of their respective Hamiltonian
is very high at $B_x=0$ and is lifted slowly (certain eigenvalues touch and do
not cross as $B_x\to0$).  Therefore, the finite values of $\gamma$ used in the
plots are not much smaller than all energy differences and we are not in the
weak dissipation limit.  As $\gamma$ is reduced, additional peaks are resolved
(cf. Fig.~\ref{openheisenberg}).

\subsection{Steady-state behavior for Ising Hamiltonians}
To get a better insight into how the condition given by
Eqn.~(\ref{conend1}) explains the peaks seen in
Fig.~\ref{6spinsising}, we now specialize to the Ising model under
collective dissipation given by Eqn.~(\ref{eqncoll}). Then we see
that the steady state apart from the degeneracy points and the
condition for discontinuity become very simple. For a detailed
derivation of what follows, see Appendix \ref{Apc}.

The Hamiltonian in Eqn.~(\ref{eqn1aneu}) with periodic boundary
conditions is in general degenerate due to translational and
reflection symmetry. To obtain a non-degenerate $H$, we restrict our
consideration to a specific subspace with eigenvalue $1$ for the
translation operator $T$ and the reflection operator $R$
\footnote{If initialized in this subspace (e.g., by optically pumping it to
  the fully polarized states $\ket{0}^{\otimes N}$), the system will remain
there since both ${\cal L}_0$ and ${\cal L}_1$ respect these
symmetries.}. Note that the Hamiltonian is also symmetric under the
spin-flip operation $F=\sigma_x^{\otimes N}$, i.e.,
$FHF^{\dagger}=H$. Using the properties of ${\cal L}_1$ and
$F$-invariance of $H$, we find that if the system has a unique steady state of ${\mathcal{L}
 }(x)$, it is, in the limit of weak dissipation, given by the maximally mixed state
$\propto\id$: plugging $\id$ into
Eqn.~(\ref{eqn6a}) we obtain $\mathrm{P}^D\mathcal{L}_1(\id)=\sum_i
\ketbra{\lambda_i(x)}{\lambda_i(x)}J^z\ketbra{\lambda_i(x)}{\lambda_i(x)}$
and flip invariance of $H$ implies
$\bra{\lambda_i(x)}J^z\ket{\lambda_i(x)}=0$ for the eigenstates of a
non-degenerate Hamiltonian $H$ (see Appendix \ref{Apc}).

Thus if the steady state is unique, it is always maximally mixed outside degeneracy
points and we see a discontinuity at $x=x_0$ if for the degenerate
eigenstates $\ket{\lambda_1(x_0)},\ket{\lambda_2(x_0)}$  we have
\begin{equation}\label{cona}
\bra{\lambda_1}J^z \ket{\lambda_2}\neq 0.
\end{equation}
This can be checked to hold for the points at which peaks are
observed in Fig.~\ref{6spinsising}.

For the Ising model in a transverse magnetic field, for larger $N$ the peaks
decrease in height, and disappear in the limit $N\longrightarrow\infty$. The
spectrum for the Ising model in a transverse field is known analytically
\cite{Pfeu70}. For large $N$, the spectrum is very dense and degeneracy points
are so closely spaced that peaks are no longer resolvable (and vanish in the
thermodynamic limit as bands develop). Nevertheless, for small spin systems,
these features provide a method to dissipatively study degeneracies of the
applicable Hamiltonian -- anywhere in the spectrum, not just in the ground
state. To the extent that $\mathcal{L}_1$ is tunable, it even provides access
to the nature of the degenerate states via Eqn.~(\ref{conend1}).

\section{Conclusions}
We have shown that using cold atoms in an optical lattice in the
Mott-insulator regime, dissipative spin chains with Hamiltonians such as the
XXZ model, the Ising model or the Heisenberg model can be realized.  Optical
driving of internal atomic states allows for the realization of a tunable
transversal magnetic field and engineered
dissipation. 

This system is an ideal testbed for studying steady-state dynamics of
dissipative spin models. We have discovered a peculiar feature of the steady
state diagram for small spin chains: in the limit of weak dissipation, the
expectation values of the collective spin operators exhibit abrupt changes
that hint at discontinuities in the steady state. These discontinuities occur
at degeneracy points of the Hamiltonian. We have studied this phenomenon for
different spin models with open and periodic boundary conditions subject to
individual and collective dissipation. Finally, we have presented conditions
that elucidate the discontinuous behavior of the steady state at degeneracy
points of the Hamiltonian. Therefore, measurements of the steady state
dynamics of cold atoms in optical lattices would allow to draw conclusions on
the spectrum of the respective spin model.

\begin{acknowledgements}
  HS and GG thank L. Mazza for useful discussions on cold
  atoms. The authors gratefully acknowledge funding by the DFG within the SFB
  631 and by the EU within project MALICIA under FET-Open grant number 265522.
\end{acknowledgements}
\appendix

\section{Derivation of effective dissipative master equation}\label{Apa}
The internal levels of the atom that we consider are $\ket{g}$,
$\ket{r}$, $\ket{e}$. Adiabatically eliminating the excited state
$\ket{e}$ as discussed in Section \ref{Diss}, we get an effective
two-level system $\ket{g}$ and $\ket{r}$ that is coupled with the
effective Rabi frequency $\Omega_{\textrm{eff}}$ as depicted in
Fig.~\ref{effectivetwolevel}.

In the following, we derive in detail the master equation given by
Eqn.~(\ref{Master}) in Section \ref{Diss}. The internal levels of
the atom that we consider are $\ket{g}$, $\ket{r}$, $\ket{e}$, as
depicted in the upper part of Fig.~\ref{effectivetwolevel}. The
states $\ket{g}$-$\ket{r}$ are coupled by a detuned Raman transition
via the excited state $\ket{e}$ by two standing wave laser fields.
The coupling is described by the Hamiltonians
\begin{equation}
H_{l1}=\sum_k\Omega_1 \cos{(k_1x_k)}(\ketbra{e}{g}_k+\textrm{h.c.}),
\end{equation}
and
\begin{equation}
H_{l2}=\sum_k\Omega_2 \sin{(k_2x_k)}(\ketbra{r}{g}_k+\textrm{h.c.}),
\end{equation}
where $\Omega_1$ and $\Omega_2$ are the Rabi frequencies of the two lasers and
$k_1$ and $k_2$ are the wave numbers of the lasers and $k$ denotes the lattice
site. $x_k$ is the displacement from the equilibrium position $x_k^0$ of the
atom at lattice site $k$. The phase of the lasers is for simplicity chosen
such that $\cos{[k_{1}(x_k+x_k^0)]}=\cos{(k_{1}x_k)}$ and
$\cos{[k_{2}(x_k+x_k^0)]}=\sin{(k_{2}x_k)}$. Choosing different phases of the
lasers makes $A^{\pm}$ in Eqn.~(\ref{Apm}) dependent on the lattice site
$k$. Adiabatic elimination of the excited state $\ket{e}$ leads to an
effective coupling
\begin{equation}\label{H1}
H_{1}=\sum_k\Omega_{\textrm{eff}}\eta_{1}(\sigma_k^-+\sigma_k^+)(\ketbra{r}{g}+\textrm{h.c.}),
\end{equation}
with
$\Omega_{\textrm{eff}}=\Omega_1\Omega_2/\delta_{re}$ where $\delta_{re}$ is
the detuning with respect to $\ket{e}$ and $\eta_{1}$ is the Lamb-Dicke parameter. Here, we have expressed the deviation
from equilibrium position, $x_k$, in terms of harmonic oscillator
operators truncated to the two lowest lying levels $\sin(k_{1}x_k)\approx\eta_{1}(\sigma_k^-+\sigma_k^+)$
where $\sigma_k^+=\ketbra{1}{0}_k$ and $\sigma_k^-=\ketbra{0}{1}_k$
and $\cos{(k_{2}x_k)}\approx 1$. The effective coupling with Rabi
frequency $\Omega_{\textrm{eff}}$ between states $\ket{r}$ and
$\ket{g}$ is shown in Fig.~\ref{effectivetwolevel}.

Coupling the state $\ket{r}$ to the excited state $\ket{e}$ with a
third standing wave laser field with Rabi frequency $\Omega_{er}$,
depicted with a red arrow in Fig.~\ref{effectivetwolevel}, we can
derive an effective two-level system $\ket{g}$-$\ket{r}$ with
designable decay rates as done in \cite{MaCi94}. Here, we briefly
review this result. Following \cite{MaCi94}, the upper level
$\ket{e}$ can be adiabatically eliminated if the saturation
parameter for the transition $\ket{r}$ and $\ket{e}$ is small
\begin{equation}
s_{r,e}=\frac{\left(\Omega_{re}/2\right)^2}{\delta_{re}^2+(\Gamma_{er}+\Gamma_{eg})^2/4}\ll
1.
\end{equation}
According to \cite{MaCi94}, the effective detuning and the effective
decay rates are given by:
\begin{equation}\label{deltar}
\delta_r=\delta_{gr}-\delta_{re}\frac{(\Omega_{re}/2)^2}{[(\Gamma_{eg}+\Gamma_{er})/2]^2+\delta_{re}^2},
\end{equation}
\begin{equation}
\Gamma=\frac{(\Omega_{re}/2)^2}{[(\Gamma_{eg}+\Gamma_{er})/2]^2+\delta_{re}^2}\Gamma_{eg},
\end{equation}
\begin{equation}
\gamma=\frac{(\Omega_{re}/2)^2}{[(\Gamma_{eg}+\Gamma_{er})/2]^2+\delta_{re}^2}\frac{\Gamma_{eg}+\Gamma_{er}}{2},
\end{equation}
see also the lower part of Fig.~\ref{effectivetwolevel}.
The effective two-level system $\ket{g}$-$\ket{r}$ with the
effective decay rates $\Gamma$, $\gamma$ and the effective detuning
$\delta_r$ is the starting point of the following discussion. The
full Hamiltonian describing the system is given by
\begin{eqnarray}\label{V3a}
H_{\textrm{full}}=H_1+H_{0},
\end{eqnarray}
where $H_{1}$ describes the atom-light interaction given by
Eqn.~(\ref{H1}) and $H_0$ defines the energies of the system
\begin{eqnarray}
H_{0}=&\sum_k\delta_r\ketbra{r}{r}_k+\nu\ketbra{1}{1}_k.
\end{eqnarray}

The effective dynamics of the system can be derived considering
contributions to the Liouvillian up to second order in a
perturbative approach. The full system is described by a Liouvillian
given by:
\begin{equation}\label{La}
\dot{\rho}(t)=(\mathcal{L}_0+\mathcal{L}_{1})\rho(t),
\end{equation}
where $\mathcal{L}_0$ is given by 
\begin{eqnarray}\label{L0a}
\mathcal{L}_{0}\rho(t)=&\sum_k\Gamma\left(2\ketbra{g}{r}_k\rho(t)\ketbra{r}{g}_k-\left\lbrace\ketbra{r}{r}_k,\rho(t)\right\rbrace_+\right)\notag\\
&+\gamma\left(2\ketbra{r}{r}_k\rho(t)\ketbra{r}{r}_k-\left\lbrace\ketbra{r}{r}_k,\rho(t)\right\rbrace_+\right)\notag\\&
-i\left[ H_{0},\rho(t)\right].
\end{eqnarray}
The first part of the Liouvillian is the decay part with the
effective decay rate $\Gamma$ from state $\ket{r}$ to $\ket{g}$ and
the dephasing rate $\gamma$. The projector 
\begin{equation}
  \label{eq:Pg}
P_g = \ketbra{g}{g}\otimes(\ketbra{0}{0}+\ketbra{1}{1})
\end{equation}
is stationary under $\mathcal{L}_0$. The perturbative part of the Liouvillian is
given by
\begin{eqnarray}\label{L1a}
\mathcal{L}_1\rho(t)=-i\left[ H_1,\rho(t)\right],
\end{eqnarray}
where $H_1$ is given by Eqn.~(\ref{H1}) and describes the interaction of the
two-level system with the effective laser field.  Treating $\mathcal{L}_1$ 
as a perturbation, we derive an effective Liouvillian in the stationary
subspace of $\mathcal{L}_0$. The projection onto this subspace reads
\begin{eqnarray}\label{Paa}
\mathbb{P}\dot{\rho}(t)=\mathbb{P}\mathcal{L}\mathbb{P}\rho(t)+\mathbb{P}\mathcal{L}\mathbb{Q}\rho(t),
\end{eqnarray}
where
$\mathbb{P}\rho=\ketbra{g}{g}\otimes(\ketbra{0}{0}+\ketbra{1}{1})\rho\ketbra{g}{g}\otimes(\ketbra{0}{0}+\ketbra{1}{1})$
and $\mathbb{Q}=1-\mathbb{P}$. Projecting onto the subspace we want
to eliminate we get
\begin{eqnarray}\label{Qaa}
\mathbb{Q}\dot{\rho}(t)=\mathbb{Q}\mathcal{L}\rho(t).
\end{eqnarray}
In the following, we integrate Eqn.~(\ref{Qaa}) to get the time
evolution of the density matrix in the fast space,
$\mathbb{Q}\rho(t)$. We insert the result in Eqn.~(\ref{Paa}) to get
an equation of motion for the density matrix in the slow space.
Therefore, we first go into the interaction picture, where the
density matrix is given by $\tilde{\rho}(t)=e^{-\mathcal{L}_0
t}\rho(t)$. The equation of motion in the fast space reads
\begin{equation}
\mathbb{Q}\dot{\tilde{\rho}}(t)=\mathbb{Q}W_I(t)\tilde{\rho}(t),
\end{equation}
with $W_I(t)=e^{\mathcal{L}_0 t}\mathcal{L}_1 e^{\mathcal{L}_0 t}$.
Solving this equation by iteration \cite{Messiah2} we get
\begin{align}\label{pandq2}
\mathbb{Q}\rho(t)&=\mathbb{Q}e^{\mathcal{L}_0t}\left[\int_0^tdsW_I(s)\mathbb{P}\tilde{\rho}(0)\right.\notag\\
&\left.+\int_0^tds_1\int_0^{s_1}ds_2W_I(s_1)W_I(s_2)\mathbb{P}\tilde{\rho}(0)\right].
\end{align}
At time $t=0$, $\tilde{\rho}(0)$=$\rho(0)$ and we assume that at
$t=0$, the population is in the ground state, i.e.,
$\tilde{\rho}(0)=\mathbb{P}\tilde{\rho}(0)$. Higher order integrals
are neglected making the assumption that
\begin{equation}\label{cond1}
\left|\Omega_{\textrm{eff}}\right| \ll \Gamma, \gamma, |\delta_r|,\nu.
\end{equation}
We denote the first integral in Eqn.~(\ref{pandq2}) by $R_1(t)$ and
the second integral by $R_2(t)$ such that
\begin{eqnarray}\label{pandq2b}
\mathbb{Q}\rho(t)=R_1(t)+R_2(t).
\end{eqnarray}
Inserting in Eqn.~(\ref{Paa}) leads to
\begin{eqnarray}\label{pandq2c}
&\mathbb{P}\dot{\rho}(t)=\mathbb{P}\mathcal{L}\mathbb{P}\rho(t)+\mathbb{P}\mathcal{L}_0R_1(t)+\mathbb{P}\mathcal{L}_1R_1(t)\notag\\&+\mathbb{P}\mathcal{L}_0R_2(t)+\mathbb{P}\mathcal{L}_1R_2(t).
\end{eqnarray}
The term $\mathbb{P}\mathcal{L}_0R_1(t)=0$, and
$\mathbb{P}\mathcal{L}_1R_2(t)$ is a third order term and can be
neglected. Neglecting terms rotating with $\exp(\pm i\nu t)$ we get
the master equation given by Eqn.~(\ref{Master}) with AC Stark shifts
given by
\begin{equation}
H_S=s_-\sigma_k^+\sigma_k^-+s_+\sigma_k^-\sigma_k^+,
\end{equation}
where 
$$s_{\pm}=\Omega_{\textrm{eff}}^2\eta_1^2\frac{(\delta_r\pm\nu)}{(\Gamma+\gamma)^2+(\delta_r\pm\nu)^2}.$$

\section{Derivation of the spin Hamiltonian}\label{Apb} 
In the Mott-Insulator regime, bosonic atoms
trapped by a lattice potential with two motional states are described
by the two-band Bose-Hubbard model
\begin{equation}\label{V1b}
H_{\textrm{BH}}=H_{0}+H_t.
\end{equation}
Here, the sum runs over the $N$ sites $k$ of the optical lattice.
The unperturbed Hamiltonian $H_{0}$ is given by
\begin{eqnarray}\label{H0bhb}
H_{0}=&\sum_k\left(\frac{U_{01}}{2}
\hat{n}_{k0}\hat{n}_{k1}+\sum_{x={0,1}}\frac{U_{xx}}{2}
\hat{n}_{kx}(\hat{n}_{kx}-1)\right.\notag\\&
\left.\phantom{\sum_{x={0,1}}\frac{U_x}{2}}+\nu\hat{n}_{k1}
\right),\notag
\end{eqnarray}
where $U_{xx'}$ is the on-site repulsion of two atoms on lattice
site $k$, where one atom is in motional state $\ket{x}$ and the
other one in $\ket{x'}$ with $x,x'=0,1$, respectively. The operator
$\hat{n}_{kx}=\ketbra{x}{x}_k$ counts the number of atoms at lattice
site $k$ in the motional states $x=0,1$ and $\nu$ is the energy
difference between ground and first excited motional states. We
assume the system to be prepared in the ground state $\ket{0}$. Due
to the anharmonicity of the potential, we do not leave the subspace
of $n=0$ and $n=1$ excitations.

The perturbative part of the Hamiltonian describes the tunneling
between neighboring lattice sites and is given by
\begin{equation}
H_t=\sum_kt_0 c_{k,0}^{\dagger}c_{k+1,0}+t_1
c_{k,1}^{\dagger}c_{k+1,1} +\mathrm{h.c.}
\end{equation}
Here, the operators $c_{kx}$ with $x=0,1$ are bosonic destruction
operators for atoms in the two motional states $\ket{0}$ and
$\ket{1}$ at lattice site $k$. $t_{0} (t_{1})$ are the tunneling
amplitudes from state $\ket{0}$ ($\ket{1}$) at lattice site $k$ to
state $\ket{0}$ ($\ket{1}$) at $k+1$.

As the on-site interaction $U_{xx'}\gg t_0,t_1$, tunneling between
neighboring wells that leads to states with two atoms in one lattice
well can be treated as a perturbation. For that, we consider two
neighboring lattice sites $k$ and $k+1$ and write the effective
Hamiltonian in the basis of eigenvectors of $H_0$,
$\ket{x_k,y_{k+1}}$, where for example $\ket{0_k,1_{k+1}}$ is the
notation for the state with one particle in well $k$ in state
$\ket{0}$, and one particle in well $k+1$ in state $\ket{1}$. In
perturbation theory \cite{CDG92}, the second-order effective
Hamiltonian can be evaluated in the following way:
\begin{eqnarray}\label{eqn10b}
\bra{x_k,y_{k+1}}H_{\textrm{\textrm{eff}}}^{(3)}\ket{x'_k,y'_{k+1}}=&\\
&\hspace*{-3cm}\frac{1}{2}\sum_{\chi}\bra{x_k,y_{k+1}}H_{t}\ket{\chi}\frac{1}{E'}\bra{\chi}H_{t}\ket{x'_k,y'_{k+1}}\notag.
\end{eqnarray}
where
$$\frac{1}{E'}=\frac{1}{E_{xy}-E_{\chi}}+\frac{1}{E_{x'y'}-E_{\chi}},$$
and $\ket{\chi}$ are eigenstates of $H_0$ with two particles in
one well (and no particle in the other
one). $E_{xy}=\bra{x_k,y_{k+1}}H_0\ket{x_k,y_{k+1}}$ and
$E_{\chi}=\bra{\chi}H_0\ket{\chi}$ are the unperturbed energies. 
Evaluating Eqn.~(\ref{eqn10b}) leads to the effective spin
Hamiltonian $H_{\textrm{eff}}^{(3)}$ given by:
\begin{eqnarray}\label{Heff3a}
H_{\textrm{eff}}^{(3)}=H_{\textrm{spin}}+B_z\sum_k\ketbra{1}{1}_k,
\end{eqnarray}
with
\begin{eqnarray}\label{Hs}
H_{\textrm{spin}}=\sum_{k}\alpha_1(\sigma_k^x\sigma_{k+1}^x+\sigma_k^y\sigma_{k+1}^y)+\alpha_2
\sigma_k^z\sigma_{k+1}^z.
\end{eqnarray}
Here, $$\alpha_1=-\frac{4t_0t_1}{U_{01}},$$
$$\alpha_2=2\left(\frac{t_0^2+t_1^2}{U_{01}}-\frac{t_0^2}{U_{00}}-\frac{t_1^2}{U_{11}}\right),$$
and $B_z$, the magnetic
field in $z$-direction is $$B_z=\frac{t_0^2}{U_{00}}-\frac{t_1^2}{U_{11}}.$$
Thus, we have derived an effective XXZ-spin Hamiltonian with a
magnetic field in $z$-direction.

\section{Condition for discontinuous behavior}\label{Apc}
Here, we first derive a general condition for the discontinuous
behavior in the steady state at a degeneracy point of a large class
of spin Hamiltonians. Then, we focus on more specific Hamiltonians.
We study the steady state of flip-invariant Hamiltonians outside the
degeneracy point and, starting with the general condition for
finding discontinuities in the steady state, we derive a more
precise condition for flip-invariant Hamiltonians.

\subsection{General condition for discontinuities in steady state}
Here, we derive a general condition for discontinuous behavior in
the steady state at the degeneracy point $x=x_0$ of a general
Hamiltonian $H$, where $H=H(x)$  is an analytic function of $x$. We
consider a system described by the master equation
\begin{equation}\label{eqn41-app}
\dot{\rho}(t)=(\mathcal{L}_0+\mathcal{L}_1)\rho(t),
\end{equation}
 where the Hamiltonian
part of the Liouvillian is given by
$\mathcal{L}_0(x)=\mathcal{L}_0=-i[H(x),\cdot]$ and depends on 
a parameter $x$, and the local decay Liouvillian is
\begin{equation}\label{eqn5a}
\mathcal{L}_1\rho(t)=\sum_k\gamma_k\left[2\sigma_k^-\rho(t)\sigma_k^+
-\left\{\sigma_k^+\sigma_k^-,\rho(t)\right\}_+\right].
\end{equation}
First, we want to describe the system outside the degeneracy point, i.e., for
$x\neq x_0$. We assume that in the vicinity of $x_0$, the Hamiltonian is
nondegenerate (for $x\neq x_0$) and that the dissipation is weak. The steady
state $\rho_{ss}(x)$ defined by
$(\mathcal{L}_0(x)+\mathcal{L}_1)\rho_{ss}(x)=0$ is, 
in the limit $\gamma\rightarrow0$ given by
\begin{equation}\label{eqn6}
\mathrm{P}^D(x)\mathcal{L}_1\mathrm{P}^D(x)\rho_{ss}=0,
\end{equation}
where $\mathrm{P}^D(x)$ is the projector onto the
$\mathrm{kernel}(\mathcal{L}_0)$. As the kernel of $\mathcal{L}_0$ is spanned by the eigenprojectors $\ketbra{\lambda_i(x)}{\lambda_i(x)}$ of $H$ we have for arbitrary $A$:
\begin{equation}\label{eqn6P-app}
\mathrm{P}^DA=\sum_i\ketbra{\lambda_i(x)}{\lambda_i(x)}A
\ketbra{\lambda_i(x)}{\lambda_i(x)},
\end{equation}
where $\ket{\lambda_i(x)}$ are
eigenstates of the Hamiltonian $H(x)$ which is assumed to be nondegenerate.

Now, let us consider the case that at $x=x_0$, the Hamiltonian has a
degeneracy point at which two or more eigenvalues cross. At this
degeneracy point, we expect an discontinuous behavior of the steady
state that leads to the peaks we observe in our numerical simulation
(see Figs. \ref{4spinsxxynichtgener}-\ref{openheisenberg}). At
$x=x_0$ the projector onto the kernel of $\mathcal{L}_0$ has to be
extended. It now  also projects onto coherences between eigenstates
of $H$: $\ket{\lambda_1}$,$\ket{\lambda_2}$ which are eigenvectors
to the degenerate eigenvalues $\lambda_1=\lambda_2$. Therefore the
projector on the coherences reads:
\begin{equation}\label{projdelta}
\mathrm{P}^{\Delta}A=\ketbra{\lambda_1}{\lambda_1}A
\ketbra{\lambda_2}{\lambda_2}+\textrm{h.c.}
\end{equation}
It is convenient to define a continuous extension of the projector
$\mathrm{P}^D$  at $x=x_0$ which reads
\begin{equation}
\mathrm{P}^D(x_0)=\lim\limits_{x \rightarrow x_0}\mathrm{P}^D(x).
\end{equation}
Thus, at $x=x_0$, the full projector onto the kernel of $\mathcal{L}_0$ reads
$\mathrm{P}^D(x_0)+P^{\Delta}$. Now the condition for the steady state
$\rho_{ss}(x=x_0)$ at the degeneracy point is given by
\begin{equation}\label{con1}
\left[\mathrm{P}^D(x_0)+\mathrm{P}^{\Delta}\right]\mathcal{L}_1\left[\mathrm{P}^D(x_0)+\mathrm{P}^{\Delta}\right]\rho_{ss}(x_0)=0.
\end{equation}
We want to find a sufficient condition for the steady state to change
discontinuously. This means that
\begin{equation}\label{con2}
\rho_{ss}(x_0)-\lim\limits_{x \rightarrow x_0}\rho_{ss}(x)\neq0,
\end{equation}
where $\lim\limits_{x \rightarrow x_0}\rho_{ss}(x)$ is the
continuous extension of $\rho_{ss}(x) \forall x\neq x_0$ to $x=x_0$.

A discontinuity in the steady state as described by Eqn.~(\ref{con2}) can
occur only if 
\begin{equation}\label{con1b}
\left[\mathrm{P}^D(x_0)+\mathrm{P}^{\Delta}\right]\mathcal{L}_1\left[\mathrm{P}^D(x_0)+\mathrm{P}^{\Delta}\right]\lim\limits_{x
\rightarrow x_0}\rho_{ss}(x)\neq0,
\end{equation}
holds, since otherwise the continuous extension $\lim\limits_{x \rightarrow
  x_0}\rho_{ss}(x)$ would be a steady state as well. 
The last part of Eqn.~(\ref{con1b}) can be simplified using
$$\left[\mathrm{P}^D(x_0)+\mathrm{P}^{\Delta}\right]\lim\limits_{x \rightarrow
  x_0}\rho_{ss}(x)=\lim\limits_{x \rightarrow x_0}\rho_{ss}(x),$$ which holds
since $\lim\limits_{x \rightarrow x_0}\rho_{ss}(x)$ is per definition in the
space onto which $\mathrm{P}^{D}(x_0)$ projects and $P^\Delta$ is orthogonal
to that space.  By Eqn.~(\ref{eqn6}) we then see that Eqn.~(\ref{con1b})
reduces to the condition
\begin{equation}\label{conend}
\mathrm{P}^{\Delta}\mathcal{L}_1\lim\limits_{x \rightarrow
x_0}\rho_{ss}(x)\neq0.
\end{equation}

If this condition is fulfilled, then $\rho_{ss}(x_0)-\lim\limits_{x
\rightarrow x_0}\rho_{ss}(x)\neq0$ which means that the steady state
shows discontinuous behavior at the degeneracy point $x=x_0$.

\subsection{Condition for discontinuous behavior for Ising Hamiltonians}

Here, we want to get a better insight how the condition given by
Eqn.~(\ref{conend}) relates to the peaks observed in our numerical
simulation. In the following, we will apply it to the Ising model in a
transverse magnetic field. In the numerical simulation (see
Fig. \ref{6spinsising}) for the Ising Hamiltonian with periodic boundary
conditions under collective dissipation described by Eqn.~(\ref{eqncoll}), we
restrict our consideration to a specific subspace with eigenvalue $1$ for
the translation operator $T$ and the reflection operator $R$: $T=R=1$. First, we want
to prove that if the steady state is unique, it is the fully mixed state
outside the degeneracy points as indicated by our numerical simulation.  Then,
we show that starting from the condition given by Eqn.~(\ref{conend}),
specialization to the Ising model allows to derive a more precise condition
for finding a discontinuity in the steady state at the degeneracy points.

First, we show that $\id$ satisfies
$(\mathcal{L}_0+\mathcal{L}_1)\id=0$ outside the degeneracy point $x\neq
x_0$. Therefore, for systems with unique steady state, it is given by the
fully mixed state for $x\neq x_0$ in the limit $\gamma\rightarrow 0$. The
Hamiltonian given by Eqn.~(\ref{eqn1aneu}) is assumed to be non-degenerate for
$x\neq x_0$ and invariant under the spin flip operator $F=\sigma_x^{\otimes
  N}$, i.e., $FHF^{\dagger}=H$. Thus, we want to show that for $x\neq x_0$:
\begin{equation}\label{eqn11b}
\mathrm{P}^D\mathcal{L}_1\id=0,
\end{equation}
where $P^D$ is given by Eqn.~(\ref{eqn6P-app}). Then,
\begin{equation}\label{eqn11c}
\mathrm{P}^D\mathcal{L}_1(\id)=2\gamma(J^-J^+ -J^+J^-)\propto
\gamma J^z.
\end{equation}
Therefore, Eqn.~(\ref{eqn11b}) reads
\begin{eqnarray}\label{11d}
  \mathrm{P}^D\mathcal{L}_1(\id)=\mathrm{P}^DJ^z=
  \sum_i
  \ketbra{\lambda_i}{\lambda_i}J^z\ketbra{\lambda_i}{\lambda_i}=0.
\end{eqnarray}
If we can show that 
Eqn.~(\ref{11d}),
\begin{equation}\label{11e}
\bra{\lambda_i}J^z\ket{\lambda_i}=0\,\forall i,
\end{equation}
then we have shown that the fully mixed state is a steady state of our system
outside the degeneracy points of the Hamiltonian. As the Hamiltonian is
nondegenerate and invariant under the flip operator $F$, the eigenvectors
  of $H$ are eigenvectors of $F$: $F\ket{\lambda_i}=\alpha_i\ket{\lambda_i}$.
  Let $\ket{\alpha}$ denote an arbitrary eigenvector of $H$ with
  $F$-eigenvalue $\alpha$. Since the spectrum of $F$ is$\{\pm1\}$, we have
  $\ket{\alpha}=\alpha^2\ket{\alpha}=\alpha F \ket{\alpha}$. Moreover, the
  flip $F$ changes the sign of $J^z$, i.e, $J^z$ and $F$ anticommute:
  $\{F,J^z\}^+=0$.  Therefore, we can write Eqn.~(\ref{11e}) as
\begin{eqnarray}\label{11f}
&\bra{\alpha}J^z\ket{\alpha}=\alpha\bra{\alpha}J^zF\ket{\alpha}
\notag\\&=-\alpha\bra{\alpha}FJ^z\ket{\alpha}=-\bra{\alpha}J^z\ket{\alpha},
\end{eqnarray}
where we have used that $\alpha^2=1$. It follows that
\begin{equation}
\bra{\alpha}J^z\ket{\alpha}=0.
\end{equation}
Consequently, $\mathrm{P}^D\mathcal{L}_1(\id)=0$ and we have shown that in
the limit of weak dissipation, the steady state, if it is unique, is the fully
mixed state. For the Ising model with up to 8 atoms and collective
dissipation, we know from our numerics that the steady state is unique.

To see that the steady state shows discontinuous behavior at the
degeneracy point $x=x_0$, we need now only to show that the
fully mixed state is not the steady state of the system. Thus, we
need to show that
\begin{equation}\label{eq:IsingCond}
\mathrm{P}^{\Delta}\mathcal{L}_1(\id)=\mathrm{P}^{\Delta}J^z=\sum_{i,j, i\neq j}'\ketbra{\lambda_i}{\lambda_i}J^z
\ketbra{\lambda_j}{\lambda_j} \neq 0
\end{equation}
where $\mathrm{P}^{\Delta}$ is given by Eqn.~(\ref{projdelta}) and $\sum'$
sums over the labels of degenerate eigenvalues. Since the eigenvectors are
orthogonal, Eqn.~(\ref{eq:IsingCond}) holds if $\exists i\neq j$ such that
\begin{equation}\label{cona-app}
\bra{\lambda_i}J^z \ket{\lambda_j}\neq 0.
\end{equation}
Therefore, Eqn.~(\ref{cona-app}) gives a condition for finding
discontinuous behavior of the steady state of the Ising model in a transverse field under collective dissipation. 
Note that this derivation can be easily extended to all non-degenerate
Hamiltonians that are flip-invariant. 

%

\end{document}